\documentclass[preprint,aps,nofootinbib,showkeys,showpacs,12pt]{revtex4-1}

\usepackage{mathrsfs,amsmath,amssymb,amsfonts}
\usepackage{graphicx,graphics,epsfig,hhline,color}

\newcommand{\be}{\begin{eqnarray}}
\newcommand{\ee}{\end{eqnarray}}

\newcommand{\Z}{\mathbb{Z}}

\newcommand{\expm}{e^{-\beta\,E^+_p}}
\newcommand{\expmm}{e^{-2\beta\,E^+_p}}
\newcommand{\expmmm}{e^{-3\beta\,E^+_p}}
\newcommand{\expp}{e^{-\beta \,E^-_p}}
\newcommand{\expppp}{e^{-3\beta\,E^-_p}}
\newcommand{\exppp}{e^{-2\beta\,E^-_p}}

\newcommand{\bc}{\begin{center}}
\newcommand{\ec}{\end{center}}

\begin{document}
\title[How parameters and regularization affect the PNJL model phase diagram]
{How parameters and regularization affect the PNJL model phase diagram and
thermodynamic quantities}

\author{P. Costa$^1$, H. Hansen$^2$, M. C. Ruivo$^3$ and C. A. de Sousa$^3$}

\affiliation{$^1$ Centro de F\'{\i}sica Computacional,
Departamento de F\'{\i}sica, Universidade de Coimbra, P-3004-516 Coimbra and E.S.T.G.,
Instituto Polit\'ecnico de Leiria, Morro do Lena-Alto do Vieiro, 2411-901 Leiria,
Portugal}
\affiliation{$^2$ IPNL, Universit\'e de Lyon/Universit\'e Lyon 1, CNRS/IN2P3, 4 rue E.Fermi,
F-69622 Villeurbanne Cedex, France}
\affiliation{$^3$ Centro de F\'{\i}sica Computacional, Departamento de F\'{\i}sica,
Universidade de Coimbra, P-3004-516 Coimbra, Portugal}


\begin{abstract}
We explore the phase diagram and the critical behavior of QCD thermodynamic quantities in
the context of the so-called Polyakov--Nambu--Jona-Lasinio model. We show that this
improved field theoretical model is a successful candidate for studying the equation of
state and the critical behavior around the critical end point. We argue that a convenient
choice of the model parameters is crucial to get the correct description of isentropic
trajectories. The effects of the regularization procedure in several thermodynamic
quantities  is also analyzed. The results are compared with simple thermodynamic
expectations and lattice data.
\end{abstract}

\maketitle
\normalsize


\section{Introduction}

Quantum chromodynamics (QCD) exhibits  at zero temperature and chemical potential two
remarkable features which play an essential role  in our understanding of strong
interaction phenomena: the fundamental degrees of freedom are colorless bound states of
hadrons (quark confinement), and  chiral symmetry is dynamically broken. As it is well
known these features characterize the nonperturbative nature of the QCD vacuum. It is
expected that, at  large energy densities, the so-called QCD phase transition occurs: the
interaction becomes weaker and weaker due to asymptotic freedom
\cite{HalaszJSSV,Shuryak2006} with the formation of a new state of matter, the quark
gluon plasma (QGP), and chiral symmetry gets restored.

The study of the  QCD phase diagram in the  $(T,\mu$)-plane and the search for signatures
of the QGP have attracted an intensive investigation over the last decades. The outputs
of this research  are expected to play an important  role in our understanding of the
evolution of the early universe and of the physics of heavy-ion collisions  at the
Brookhaven National Laboratory, and at LHC (CERN),   in the future.

Various results from QCD inspired models indicate (see e.g. Refs.
\cite{Barducci:1994PRD,Stephanov:1996PRL}) that at low temperatures the transition may be
first order for large values of the chemical potential; on the contrary a crossover is
found for small chemical potential and large temperature.
This suggests that the first order transition line may end when the temperature
increases, the phase diagram thus exhibiting a (second order) critical endpoint (CEP)
\cite{Schwarz:PRC1999,Buballa:2003PLB,Costa:2007PLB,Asakawa} that can be detected
\cite{Stephanov:1998PRL,Hatta:2003PRD} by a new generation of experiments with
relativistic nuclei, as the CBM experiment  (FAIR) at GSI.
Fodor and Katz \cite{Karsch} claim the values $T^{CEP}=162$ MeV and $\mu^{CEP}=360$  MeV
for such a critical point, although its precise location is still a matter of debate
\cite{Gavai:2005PRD}.
In the chiral limit it is found, in accordance with universality arguments, a tricritical
point (TCP) in the phase diagram, separating the second order transition line from the
first order one.
So, the exploration of the critical region  of the phase diagram of strongly interacting
matter gains increasing attention, both experimentally and theoretically.

As an approach complementary to first-principle lattice simulations, one can consider
several effective models. One of them is the Nambu--Jona-Lasinio (NJL) model, that  is
undoubtedly a useful tool for understanding chiral symmetry breaking but does not possess
a confinement mechanism. As a reliable model that can treat both the chiral and the
deconfinement phase transitions, we can consider the Polyakov loop extended NJL (PNJL)
model \cite{Megias:2006PRD,Pisa1,Ratti:2005PRD,Muller}. In the PNJL model the
deconfinement phase transition is described by the Polyakov loop. This improved field
theoretical model is fundamental for interpreting the lattice QCD results and
extrapolating into regions not yet accessible to lattice simulations.

A nontrivial question in NJL type models is the choice of the parameter set and of the
regularization procedure. In fact, one should keep in mind that this type of models  are
used not only to describe physical observables in the vacuum but also to explore the
physics at finite temperature and chemical potential. As it is well known, the order of
the phase transition on the axis of the $(T,\mu)$-plane is sensitive to the values of the
parameters, most notably to the value of the ultraviolet cutoff needed to regularize the
integrals. In the pure NJL model a  large cutoff leads to a second order transition, a
small cutoff to a first order one \cite{Klevansky}. An interesting feature to be noticed
is that the requirement of global accordance with physical spectrum is obtained with
values of the cutoff for which the transition is first order on the $T=0$ axis and a
smooth crossover on the $\mu=0$ axis of the phase diagram. However, it has also been
shown  that different parameter sets, although providing  a reasonable fit of hadronic
vacuum observables and predicting a first order phase transition, will lead to different
physical scenarios at finite $T$ and $\mu$ \cite{Buballa:2004PR,Costa:2003PRC}. For
instance, the absolute stability of the vacuum state at $T=0$ is not always insured.

The consequences of the choice of the parameter set for the scenarios in the
$(T,\mu)$-plane have not been discussed in the framework of the PNJL model, where the
most popular parameter set does not allow for the absolute stability of the vacuum at
$T=0$.
In the present work, our main goal is to  analyze this problem and we will assume that
the most reliable parametrization of both NJL and PNJL models positively predicts the
existence of the CEP in the phase diagram, together with the formation of stable quark
droplets in the vacuum state at $T=0$.

Finally,  the physical relevance of our numerical solutions is insured by requiring the
agreement with general thermodynamic requirements.  In particular, we   will verify that
the correct description of isentropic lines is closely related with the parameter choice
in the pure NJL sector.

Concerning the  regularization of some integrals, since, as it has been noticed by
several authors, the three dimensional cutoff is only necessary at zero temperature, the
dropping of this cutoff at finite $T$ is carefully analyzed in this work:  this procedure
allows for the presence of high momentum quark states, leading to interesting physical
consequences, as it has been shown in \cite{Costa:2008PRD2}, where the advantages and
drawbacks of this regularization have been discussed. We will enlarge the use of this
procedure to the PNJL model and discuss its influence on the behavior of several relevant
observables.

Let us notice that the choice of a regularization procedure is a part of the
effective modeling of QCD thermodynamic. Indeed  the presence of high momentum quarks (no
cutoff on the temperature dependent part of the loop integrals) is required to ensure
that the entropy scales as $T^3$ at high temperature.
Hence we found that a
comprehensive study of the differences between the two regularization procedures (with
and without cutoff on the quark  momentum states at finite temperature) is necessary to
have a better understanding of the PNJL model and the role of high momentum quarks around
the phase transition. This is one of the main purposes of this paper.


This paper is organized as follows: In Sec. II we present the model and formalism starting
with the deduction of the self-consistent equations. We also extract the equations of
state and the response functions, and show the pertinence and physical relevance of a
convenient choice of the parametrization  and regularization procedures of the model.
Section III  is devoted to  the study of the phase transition at zero temperature, showing the
important role of the choice of parameters for the formation of quark droplets in
mechanical equilibrium with the vacuum at zero pressure. In Sec. IV we study
thermodynamical quantities that,   compared with lattice results, point out the
physically relevant regularization procedure at $T\neq 0$. The enlargement to $\mu\neq 0$
allows for the study of the phase diagram in the $(T,\mu)$-plane (Sec. V). In Sec. VI we
discuss the important role of the choice of the model parameters for the correct
description of isentropic trajectories. In Sec. VII we proceed to study the size of the
critical region around the CEP and its consequences for the susceptibilities and critical
exponents. Finally, some concluding remarks are presented in Sec. VIII.


\section{Model and formalism}


\subsection{Model Lagrangian and gap equations}

The generalized Lagrangian of the  quark model for $N_f=2$ light quarks and $N_c=3$ color
degrees of freedom,   where the quarks are coupled to a (spatially constant) temporal
background gauge field (represented in term of Polyakov loops), is given by
\cite{Pisa1,Ratti:2005PRD,Fukushima:2008PRD}:
\begin{eqnarray}
{\mathcal L_{PNJL}\,}&=& \bar q\,(\,i\, {\gamma}^{\mu}\,D_\mu\,-\,\hat m)\,q +
\frac{1}{2}\,g\,[\,{(\,\bar q\,q\,)} ^2\,\,+\,\,{(\,\bar q
\,i\,\gamma_5\,\vec{\tau}q\,)}^2\,] \nonumber\\ 
&-& \mathcal{U}\left(\Phi[A],\bar\Phi[A];T\right),
\label{eq:lag}
\end{eqnarray}
where the quark fields $q\,=\,(u,d)$ are defined in Dirac and color spaces, and
$\hat{m}=\mbox{diag}(m_u,m_d)$ is the current quark mass matrix. The pure NJL sector
contains three parameters: the coupling constant $g$, the cutoff $\Lambda$, and the
current quark mass $m=m_u=m_d$, to be determined (see Sec. II C) by fitting the experimental
values of several physical quantities.

\begin{table}[t]
\begin{center}
\begin{tabular}{cccccccccccccccc}
  \hline\hline

  $a_0$ &&& $a_1$ &&& $a_2$ &&& $a_3$ &&& $b_3$ &&& $b_4$ \\
  \hline
  6.75 &&& -1.95 &&& 2.625 &&& -7.44 &&& 0.75 &&& 7.5 \\
  \hline\hline
  \end{tabular}
	\caption{
  \label{table:paramPNJL}
  Parameter set used for the Polyakov loop potential (2) and  (3).}
\end{center}
\end{table}

The quarks are coupled to the gauge sector {\it via} the covariant derivative
$D^{\mu}=\partial^\mu-i A^\mu$. The strong coupling constant $g_{Strong}$ has been
absorbed in the definition of $A^\mu$: $A^\mu(x) = g_{Strong} {\cal
A}^\mu_a(x)\frac{\lambda_a}{2}$, where ${\cal A}^\mu_a$ is the SU$_c(3)$ gauge field and
$\lambda_a$ are the Gell--Mann matrices. Besides in the Polyakov gauge and at finite
temperature $A^\mu = \delta^{\mu}_{0}A^0 = - i \delta^{\mu}_{4}A^4$.

The Polyakov loop $\Phi$ (the order parameter of $\Z_3$ symmetric/broken phase transition
in pure gauge) is the trace of the Polyakov line defined by:
$ \Phi = \frac 1 {N_c} {\langle\langle \mathcal{P}\exp i\int_{0}^{\beta}d\tau\,
    A_4\left(\vec{x},\tau\right)\ \rangle\rangle}_\beta$,
 where ${\langle\langle  \ldots \rangle\rangle}_\beta$ with $\beta = 1/T$ is the thermal
expectation value in the grand canonical ensemble.

The pure gauge sector is described by an effective potential
$\mathcal{U}\left(\Phi[A],\bar\Phi[A];T\right)$ that takes the form

\begin{eqnarray}
    \frac{\mathcal{U}\left(\Phi,\bar\Phi;T\right)}{T^4}
    &=&-\frac{b_2\left(T\right)}{2}\bar\Phi \Phi-
    \frac{b_3}{6}\left(\Phi^3+
    {\bar\Phi}^3\right)+ \frac{b_4}{4}\left(\bar\Phi \Phi\right)^2\;,
    \label{Ueff}
\end{eqnarray}
where
\begin{eqnarray}
    b_2\left(T\right)&=&a_0+a_1\left(\frac{T_0}{T}\right)+a_2\left(\frac{T_0}{T}
    \right)^2+a_3\left(\frac{T_0}{T}\right)^3.
\end{eqnarray}
The coefficients $T_0$, $a_i$ and $b_i$ of the Polyakov loop effective potential are chosen 
(see Table \ref{table:paramPNJL} and \cite{Ratti:2005PRD}) to reproduce, at the mean-field
level, the results obtained in pure gauge lattice calculations. 

From the Lagrangian (\ref{eq:lag}) in the mean-field approximation it is
straightforward (see Ref.~\cite{Klevansky-review}) to obtain  the constituent  quark mass
$M$  that is given by

\begin{equation}
M=m-2g\,\left\langle\bar{q} q\right\rangle ,\label{eq:gap}
\end{equation}
where  the quark condensate $\left\langle\bar{q} q\right\rangle$ has to be
determined in a self-consistent way.
So, taking already into account Eq. (\ref{eq:gap}), the PNJL grand   potential density in
the mean-field approach is given by \cite{Ratti:2005PRD,Hansen:2007PRD}:

\be
\Omega(\Phi, \bar\Phi, M ; T, \mu)&=&{\cal U}\left(\Phi,\bar{\Phi},T\right) +2g\,
N_f\left\langle\bar{q}q\right\rangle^2
- 2 N_c\,N_f \int \frac{\mathrm{d}^3\,{p}}{\left(2\pi\right)^3}\,{E_p}\nonumber \\
&+&2N_f\,T\int \frac{\mathrm{d}^3\,{p}}{\left(2\pi\right)^3} \,\left[
   \ln N^+_\Phi\,(E_p)\,+\,
   \ln\,
N^-_\Phi\,(E_p)\right] , \label{omega} 
\ee
where  $E_p$ is the quasiparticle energy for the quarks,
$E_{p}=\sqrt{{\vec{p}\,}^{2}+M^{2}}$, and by  defining $ E_p^{\pm}\,=\,E_p\,\mp \mu$, the
upper sign applying for fermions and the lower sign for antifermions, the functions
$N^+_\Phi$ and $N^-_\Phi$ are:
\be N^+_\Phi(E_p) &\equiv&  \left[ 1 + 3\left( \Phi + \bar\Phi \expm \right) \expm
 + \expmmm \right]^{-1}~, \ \label{eq:termo1}\label{zplus}
\\
N^-_\Phi(E_p) &\equiv& \left[ 1 + 3\left( \bar\Phi + \Phi \expp \right) \expp
 + \expppp \right]^{-1}~.\ \label{eq:termo2}\label{zmoins}
\ee

It was shown in Ref.~\cite{Hansen:2007PRD}  that all calculations in the NJL model can be
generalized to the PNJL one by introducing the modified Fermi--Dirac distribution
functions for particles and antiparticles:
\be f^{+}_\Phi(E_p)  = M^+_\Phi \, N^+_\Phi,\hskip0.2cm \,M^+_\Phi\,=\,\left(
\Phi\,+\,2\,  \bar\Phi  \,\expm \right)\,\expm \,+\,\expmmm~, \\
f^{-}_\Phi(E_p)  = M^-_\Phi \, N^-_\Phi,  \hskip0.2cm \,M^-_\Phi\,=\,\left(
\bar\Phi\,+\,2\, \Phi  \,\expp \right)\,\expp \,+\,\expppp~. 
\label{fpPhi} 
\ee

To obtain the mean-field equations we must search for the minima of the thermodynamical
potential density (\ref{omega}) with respect to $\left\langle\bar{q} q\right\rangle$,
$\Phi$ and $\bar\Phi$. In fact, by minimizing $\Omega$ with respect to
$\left\langle\bar{q} q\right\rangle$, we obtain the equations for the quark condensate:
\begin{equation}
\left\langle\bar{q} q\right\rangle\,=\, - 6\,\int
\frac{\mathrm{d}^3\,{p}}{\left(2\pi\right)^3} \frac{M}{E_p}[\theta\,(\Lambda^2\,-\,{\vec
p}^2)\,-\,f^+_\Phi(E_p)-f^-_\Phi(E_p)]\,. \label{eq:qqbar}
\end{equation}

Let us stress that in the latter we used a cutoff $\Lambda$ only for the $T=0$ part
of the integral (denoted case I in the following); one could put a global cutoff
$\int_0^\Lambda$ (denoted case II in the above).

Furthermore,  minimization of $\Omega$ with respect to $\Phi$ and $\bar\Phi$ provides,
respectively, the two additional mean-field equations~\cite{Hansen:2007PRD} (the integral
that appears has to be understood as $\int_0^\infty$ for case I and $\int_0^\Lambda$ for
case II):
\be 
0&=& \frac{T^4}{2} \left[-b_2(T) \bar\Phi - b_3 \Phi^2 + b_4 \Phi \bar\Phi^2\right] \nonumber\\
&-& 12\, T \int \frac{\mathrm{d}^3\,{p}}{\left(2\pi\right)^3} \left [ \expmm\,N^+_
\Phi\,+\, \expp\,N^-_ \Phi \right]\,, 
\label{eq:domegadfi}
\ee
\be
0&=& \frac{T^4}{2} \left[-b_2(T) \Phi - b_3 \bar\Phi^2 + b_4 \bar\Phi \Phi^2\right] \nonumber\\ 
&-& 12\, T \int \frac{\mathrm{d}^3\,{p}}{\left(2\pi\right)^3} \left [ \expm\,N^+_ \Phi\,+\,
\exppp\,N^-_ \Phi \right]\,. 
\label{eq:domegadfi2}
\ee

The solutions of the three coupled equations (\ref{eq:gap}), (\ref{eq:domegadfi}) 
and (\ref{eq:domegadfi2}) allow us to obtain the behavior of $M$, and 
the Polyakov loop expectation values as a function of $T$ and $\mu$.


\subsection{Equations of state and response functions}

From the thermodynamical potential $\Omega(T,\mu)$ one can derive equations of state
which allow us to  compare some of our results with  observables that have become
accessible in lattice QCD at nonzero chemical potential. The relevant observables are the
(scaled) quark number density, defined as
\begin{equation}
\frac{\rho_q(T,\mu)}{T^3} = -\frac{1}{T^3}\left (\frac{\partial\Omega}{\partial\mu}\right )_T\,,
\end{equation}
and the (scaled) ``pressure difference'' given by
\begin {equation}
\frac{\Delta p(T,\mu)}{T^4}=\frac{p(T,\mu)-p(T,0)}{T^4}\, .
\end {equation}

As usual, the pressure, $p$, is defined such as its value is zero in the vacuum state
\cite{Buballa:2004PR} and, since the system is uniform, one has
\begin{equation} \label{p}
    p (T,\mu) = -  \frac{\Omega(T,\mu)}{ V}\,,
\end{equation}
where $V$ is the  volume of the system.

Our work also includes the study of the isentropic trajectories due to their relevance 
for the study of the thermodynamics of matter created in relativistic heavy-ion collisions.
The equation of state for the entropy density, $s$, is given by
\begin{equation}
s(T,\mu)\,= \left (\frac{\partial p}{\partial T}\right)_{\mu}\,,
\end{equation}
and  the  energy density, $\epsilon$, comes from the following fundamental relation of
thermodynamics

\begin{equation}\label{energydens}
    \epsilon (T, \mu)\,=\,T\,s(T,\mu)\,+\,\mu\,\rho_q(T,\mu)\,-\,p(T,\mu)\,.
\end{equation}

The energy density, as well as the pressure, is defined such that its value is zero in the vacuum state
\cite{Buballa:2004PR}.

The baryon number susceptibility $\chi_q$ and the specific heat $C$ are  the response of the
baryon number density $\rho_q(T,\mu)$ and the entropy density $s (T,\mu)$ to an infinitesimal
variation of the quark chemical potential $\mu$ and temperature, given respectively by:
\begin{equation}
    \chi_q = \left(\frac{\partial \rho_q}{\partial\mu}\right)_{T}, \hskip1cm  {\rm and}
    \hskip1cm  C = \frac{T}{V}\left ( \frac{\partial s}{\partial T}\right)_{\mu}.
    \label{chi}
\end{equation}

These second order derivatives of the pressure are relevant quantities to discuss phase
transitions, mainly the second order ones.


\subsection{Model parameters and regularization schemes}

As already stated, the pure NJL sector involves three parameters: the coupling constant
$g$, the current quark mass $m$ and the cutoff $\Lambda$. These parameters are determined
in the vacuum by fitting the experimental values of several physical quantities.
We notice that the parameters $g$ and $\Lambda$ are correlated with each other: if we
increase $g$ in order to provide a more significative attraction between quarks, we must
also increase the cutoff $\Lambda$ in order to insure a good agreement with experimental
results. In addition,  the value of the cutoff itself does have some impact as far as the
medium effects in the limit $T=0$ are concerned.

We remember that different parameterizations may  give rise to different physical
scenarios at $T=0$ and $\mu\neq 0$ \cite{Buballa:2004PR}, even if they give reasonable 
fits to hadronic vacuum observables  and predict a first order phase transition.
%
\newcommand{\quarkdensity}{$|\langle{\bar \psi}_u\psi_u\rangle|^{1/3}$}
\begin{table}[t]
    \begin{center}
        \begin{tabular}{||c||c|c|c||c|c|c|c||}
            \hhline{|t:===:t:=====:t|}
            $ $ & $\Lambda$ &  $g$   &  $m$ &\phantom{\bigg(}\quarkdensity\ & $f_{\pi}$ & $m_{\pi}$ & $M$ \\
            $ $ & [GeV] & [GeV$^{-2}$] & [MeV] & [MeV] & [MeV] & [MeV] & [MeV]\\
            \hline
            Set A$  $ &\phantom{\bigg(} 0.590 &               7.0 &           6.0 &
                   241.5 &               92.6 &         140.2 & 400\\
            \hline
            Set B$  $ &\phantom{\bigg(} 0.651 &               5.04 &           5.5 &
                   251 &               92.3 &         139.3 & 335\\
            \hhline{|b:===:b:=====:b|}
        \end{tabular}
    \caption{    
    \label{table:paramNJL}
    Set of parameters ($\Lambda,\,g,\,m$) used in the  NJL sector of the PNJL model and the
     physical quantities chosen to fix the parameters. The constituent quark mass obtained is also included.}
    \end{center}
\end{table}
%
Here, we will use two different sets of parameters whose values are presented  in Table
\ref{table:paramNJL}. These two sets are the most widely used in NJL type models:  set A is taken from
\cite{Buballa:2003PLB} and set B from \cite{kunihiro}, the last one  being commonly used
in the context of the PNJL model \cite{Ratti:2005PRD,Hansen:2007PRD}.
 The main feature  is  a lower (larger) value of the cutoff for set A (B), for which
we verify that $\Lambda/M<1.8$ ($>1.8$).
We notice that the transition between the regime of stable to the regime of
metastable quark matter occurs at the value $\Lambda/M\approx  1.8$ \cite{fiolhais}. So,
we will prove that only the set of parameters A insures the stability conditions and,
consequently, the   compatibility with  thermodynamic expectations.
We notice that set A also agrees with an empirical relation derived in
\cite{Buballa:1996NPA} which states that stable quark matter is only possible in NJL
model if $M > 4 f_\pi$.

On the other hand, the  regularization procedure, as soon as the temperature effects are
considered, has relevant consequences on the behavior of physical observables, namely on
the chiral condensates and the meson masses \cite{Costa:2008PRD3}. In PNJL model, the two
types of   regularization  may be found
\cite{Ratti:2005PRD,Hansen:2007PRD,Sasaki,Kashiwa:2008PLB,Alb2009} as well as different
sets of parameters
\cite{Ratti:2005PRD,Hansen:2007PRD,Costa:2008PRD3,Sasaki,Kashiwa:2008PLB}.
So, in order to compare the differences between the use of different sets of
parameters and of regularization schemes,  in the physical scenarios of the PNJL model,
we will consider sets A and B of parameters and two different regularization procedures
at $T\neq 0$:

{\em Case I.} {---} The cutoff is used  only in the integrals that are divergent
($\Lambda \rightarrow \infty$ in the convergent ones; see
Eqs.(\ref{eq:qqbar}, \ref{eq:domegadfi}, \ref{eq:domegadfi2}) for
example) at finite temperature, a procedure that allows us to take into account the effects
of high momentum quarks \cite{Ratti:2005PRD,Klevansky,Sasaki,Mish2000}.

{\em Case II.} {---}  The regularization consists in the use of the cutoff $\Lambda$ in
all integrals \cite{Hansen:2007PRD,Kashiwa:2008PLB}.

Advantages and drawbacks of these regularization procedures have been discussed in
\cite{Costa:2008PRD2}. Here, our main goal is to show nontrivial consequences of the
regularization scheme used in case I.  We remind that the main drawback of this
regularization is that  at high temperature there is a too fast decrease of the quark
masses that  become lower than their current values. This    leads to a non physical
behavior of the quark condensates that, after vanishing at the point where constituent
and current quark masses coincide, changes sign and acquires  a nonzero value again.
Therefore, if we want to keep calculating observables in this region, it seems sensible
to impose the condition that the quark masses take their current values and the quark
condensates remain zero. This is the approach used here.


\section{Phase transition at zero temperature}

Our study refers mainly to the finite temperature case. However, the particular
case of zero temperature is very important due to the possibility of having, simultaneously,
a vanishing density and a finite chemical potential.
This feature depends on the choice of the parameters and  is necessary in order to
insure the satisfaction of general thermodynamic requirements.

Here we analyze the stability of the quark matter  at $T=0$. For this special case the
PNJL model reduces to the NJL one. We will now present a discussion about the stability
of the system along the same lines of \cite{Buballa:2004PR,Costa:2003PRC,Costa:2008PRD1},
which will allow us to choose the set of parameters corresponding to the most convenient
physical situation.

In the limit $T\rightarrow 0$ the normal Fermi-Dirac distribution function  reduces to
the step function
\begin{equation}
\theta (\mu\,-\,E_p)\,=\,\theta (p_F-p) \,\theta (\mu\,-\,M),
\end{equation}
where  the Fermi momentum is given by
\begin{equation}\label{step}
p_F\,=\,\sqrt{\mu^2\,-\,M^2}\,\theta (\mu\,-\,M)\,=\,(\pi^2\,\rho_q)^{1/3}\,\theta
(\mu\,-\,M).
\end{equation}

The important point of our argumentation about the choice of the model parameters comes
from the comparison between  the point $(0,\mu_{c})$ of the phase diagram, where
$\mu_{c}$ is the position of the first order line at zero temperature, and the point
$(0,M_{vac})$, where $M_{vac} = M$ is the mass of the $u,d$-quark in the vacuum.
 Two special cases are observed \cite{Buballa:2004PR}:
\begin{itemize}
  \item [(i)] For set A, the first order phase transition occurs at $\mu_{c}$ such that $\mu_{c}<M_{vac}$, and
  consequently (see Eq. \ref{step}) the phase transition connects the vacuum state ($\rho_{q}=0$) directly with the
   phase of partially restored chiral symmetry ($\rho_{q}=\rho_{c}$).
  \item [(ii)]  For set B,  $\mu_{c}> M_{vac}$, so  the phase transition connects a $\rho_{q}\neq 0$  phase of
   massive quarks with  the phase of partially restored chiral symmetry ($\rho_{q}=\rho_{c}$).
\end{itemize}

So, although we can choose several sets of parameters which fit physical observables in
the vacuum, we notice, however, that the value of the cutoff itself does have some impact
on the characteristic of the first order phase transition. Comparing  the two sets of
parameters we verify that for larger values of the cutoff, as in set B, a more strong
attraction is necessary both to reproduce the physical values in the vacuum and to insure
a  first order phase transition. As we will argue in the sequel, the more reliable case
is provided by set A.

In case (i) the energy per particle  reaches, at $\rho_q=\rho_{c}$, an absolute minimum
$\epsilon < \epsilon (0)\,=\,M_{vac}$. This is compatible with the existence of stable
quark matter, indicating the possibility for finite droplets to be in mechanical
equilibrium with the vacuum at zero pressure
\cite{Buballa:2004PR,Costa:2003PRC,Mish2000,Rajagopal:1999NPA}.
This is due to the fact that the pressure has three zeros, respectively at $\rho=0,\,
0.52 \rho_0, \,4.3\rho_0$ ($\rho_0 = 0.17$ fm$^{-3}$), that correspond to extrema  of the
energy per particle. The third zero of the pressure,  located at  $\rho_{c}=4.3 \rho_0$,
corresponds to an absolute minimum of the energy.
The critical point of the phase transition in these conditions satisfies to
$\mu_{c}<M_{vac}$ \cite{Buballa:2004PR,Scavenius}. This can be seen  by comparing
$\mu_{c}=383$ MeV with the quark masses $M_{vac}\,=\,M\,=\,400$ MeV. Above
$\rho_q=\rho_{c}$, we have again a uniform gas phase. For densities $0<\rho_q<\rho_{c}$
the equilibrium configuration is a mixed phase, where the equality of all intensive
variables ($T$, $P$ and $\mu$) defines the condition for the phase equilibrium.
So, the Gibbs criterion  is satisfied and the phase transition is a first order one
\cite{Buballa:2004PR,Costa:2008PRD1}.

On the contrary, the minimum at $\rho_q=\rho_{c}$ in case (ii)  corresponds to metastable
quark matter, since $\epsilon > \epsilon (0)\,=\,M_{vac}$.
The pressure still has three zeros, respectively at $\rho_q=0,\, 1.51 \rho_0,\,
2.64\rho_0$, that correspond to extrema of the energy per particle. The main difference
now is that the absolute  minimum of the energy per particle is at $\rho_q=0$.
This means that, in spite of
being in the presence of a first order phase transition, there are no droplets in
mechanical equilibrium with the vacuum at zero pressure. In addition, from the physical
point of view, this scenario is unrealistic because it predicts the existence of a
low-density phase of homogeneously distributed constituent quarks \cite{Buballa:2004PR}.
Other implications of this scenario on the reliability of isentropic trajectories will be
discussed later.


\section{Thermodynamic quantities}

\begin{figure}[t]
\begin{center}
  \begin{tabular}{cc}
    \hspace{-0.5cm}\epsfig{file=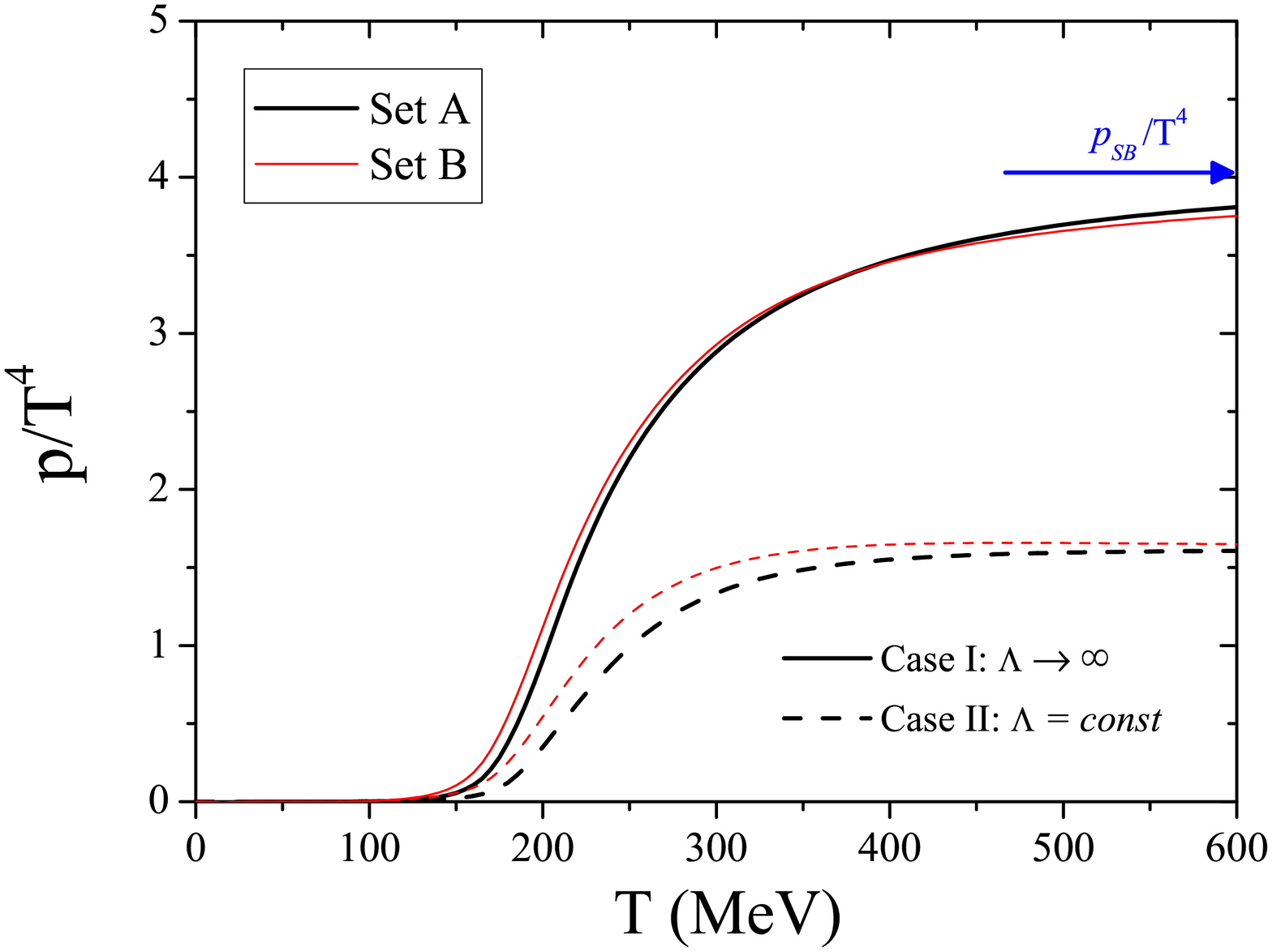,width=8.5cm,height=7cm} &
    \hspace{-0.75cm}\epsfig{file=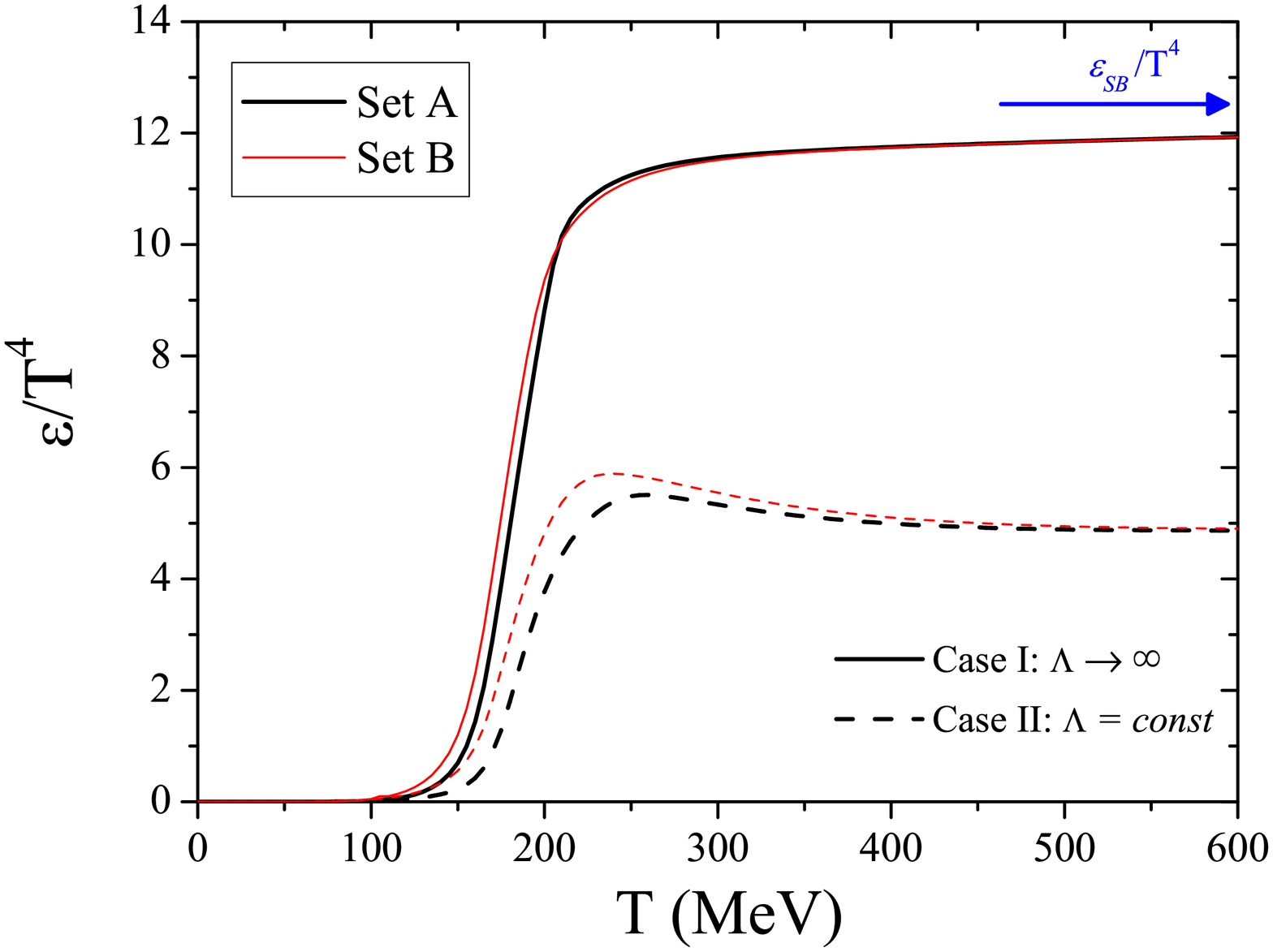,width=8.5cm,height=7cm} \\
   \end{tabular}
  \epsfig{file=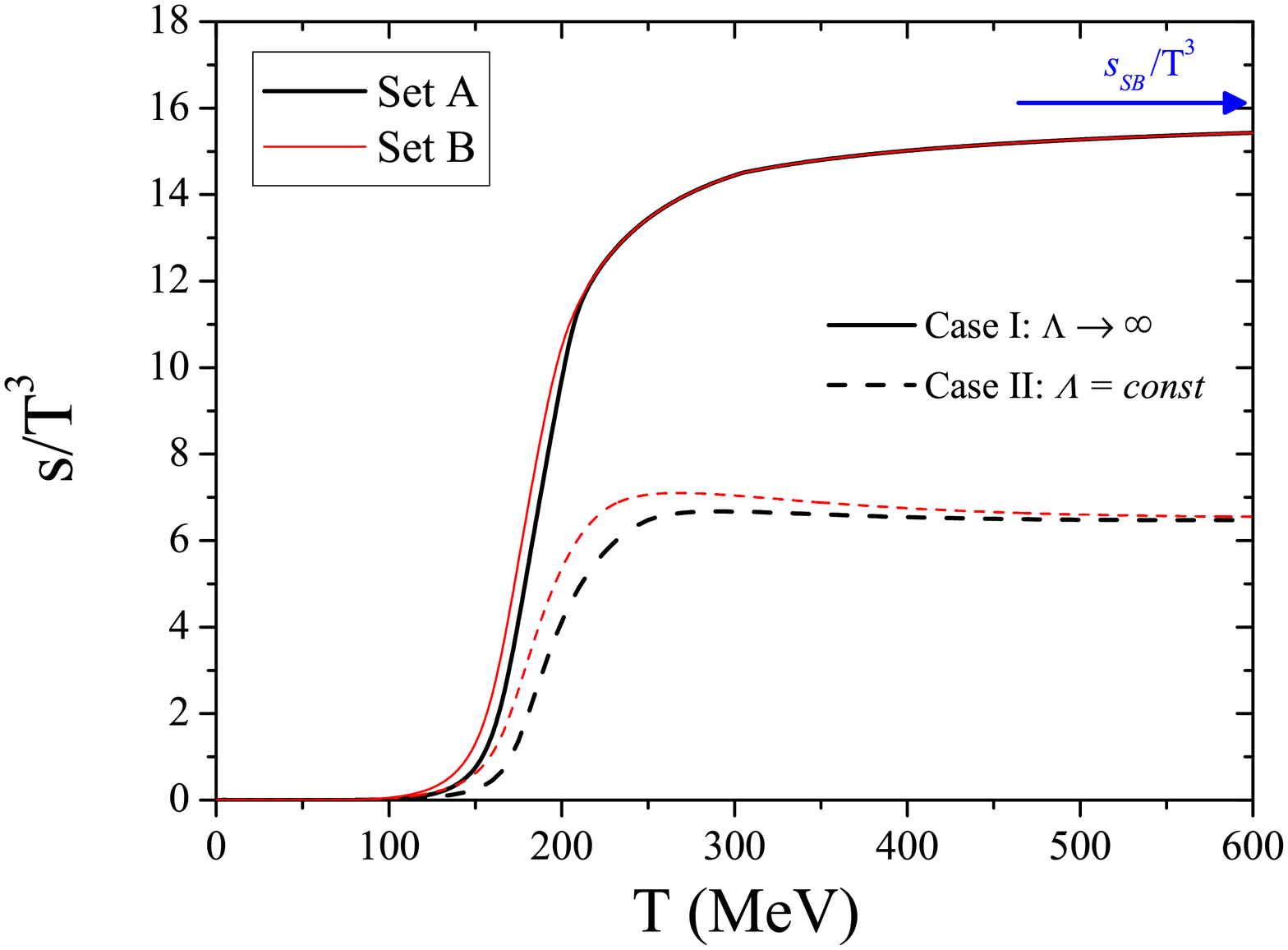,width=8.5cm,height=7cm}
\end{center}
\vspace{-1.0cm} 
\caption{ Scaled pressure $(p)$,  energy  per particle $(\epsilon)$, and
entropy $(s)$ as a function of the temperature at zero chemical potential for both sets
of parameters A and B, and both regularization procedures.} \label{Fig:1}
\end{figure}

A significative information on the phase structure of QCD at high temperature is obtained
from lattice calculations in the limit of vanishing quark chemical potential. The
transition to the phase characteristic of this regime is related with chiral and
deconfinement transitions which are the main features of our model calculation.

Following the argumentation presented in \cite{Ratti:2005PRD}, we use the reduced
temperature $T_c$ by rescaling the parameter $T_0$ from 270 to 190 MeV  (we do this
rescaling only for the remainder of this section). In this case we loose the perfect
coincidence of the chiral and deconfinement transitions: they are shifted relative to
each other by less than 35 (30) MeV for set A (B). As in Ref. \cite{Ratti:2005PRD}, we
define $T_c$ as the average of the two transition temperatures: we have $T_c = 190\,
(184)$ MeV for set A (B) within the range expected from lattice calculations
\cite{Karsch2}.

 For comparison purposes with lattice findings, we start by considering our numerical results
at vanishing quark chemical potential by checking the usefulness of the present
regularization procedure, case I. To this purpose, we plot   the scaled pressure, the
energy and the entropy as functions of the temperature in Fig. \ref{Fig:1}.

The transition to the high temperature phase is a rapid crossover rather than a phase
transition and, consequently, the pressure, the entropy and the energy densities are
continuous functions of the temperature.
 For case I we observe a similar behavior in the three curves: a sharp increase in the
vicinity of the transition temperature and then a tendency to saturate at the
corresponding ideal gas limit.
Asymptotically, the QCD pressure  for $N_f$ massless quarks and $(N_c^2 - 1)$ massless
gluons is given, for $\mu=0$,  by
\begin{equation}\label{pSB}
\frac{p_{SB}}{T^4}\,=\,(N_c^2 - 1)\,\frac{\pi^2}{45}\,+\,N_c\,N_f\,\frac{7\,\pi^2}{180},
\end{equation}
 where the first term denotes the gluonic contribution and the second term the fermionic one.

Our results follow the expected tendency and go to the free gas  values, a feature that
was also found  with this type of regularization in the context of the PNJL model
\cite{Megias:2006PRD,Ratti:2005PRD}.
For what concerns the NJL model \cite{Klevansky} let us notice that if indeed a tendency
to saturate is found, the asymptotic value is at about half the ideal gas limit.
Hence the inclusion of the Polyakov loop effective potential ${\cal U}(\Phi,\bar\Phi)$
(it can be seen as an effective pressure term mimicking the gluonic degrees of freedom of
QCD) is required to get the correct limit.

The  inclusion of the Polyakov loop together with the regularization procedure
implemented in case I, is essential to obtain the required increase of extensive
thermodynamic quantities, insuring the convergence to the Stefan-Boltzmann (SB) limit of
QCD \cite{lattice1}. Some comments are in order concerning the role of the regularization
procedure for $T>T_c$. In this temperature range, due to the presence of high momentum
quarks, the physical situation is dominated by the significative decrease of the
constituent quark masses by the $q \bar q$ interactions.  This allows for an ideal gas
behavior of almost massless quarks with the correct number of degrees of freedom.
%

\begin{figure}[t]
\begin{center}
  \begin{tabular}{cc}
    \hspace{-0.25cm}\epsfig{file=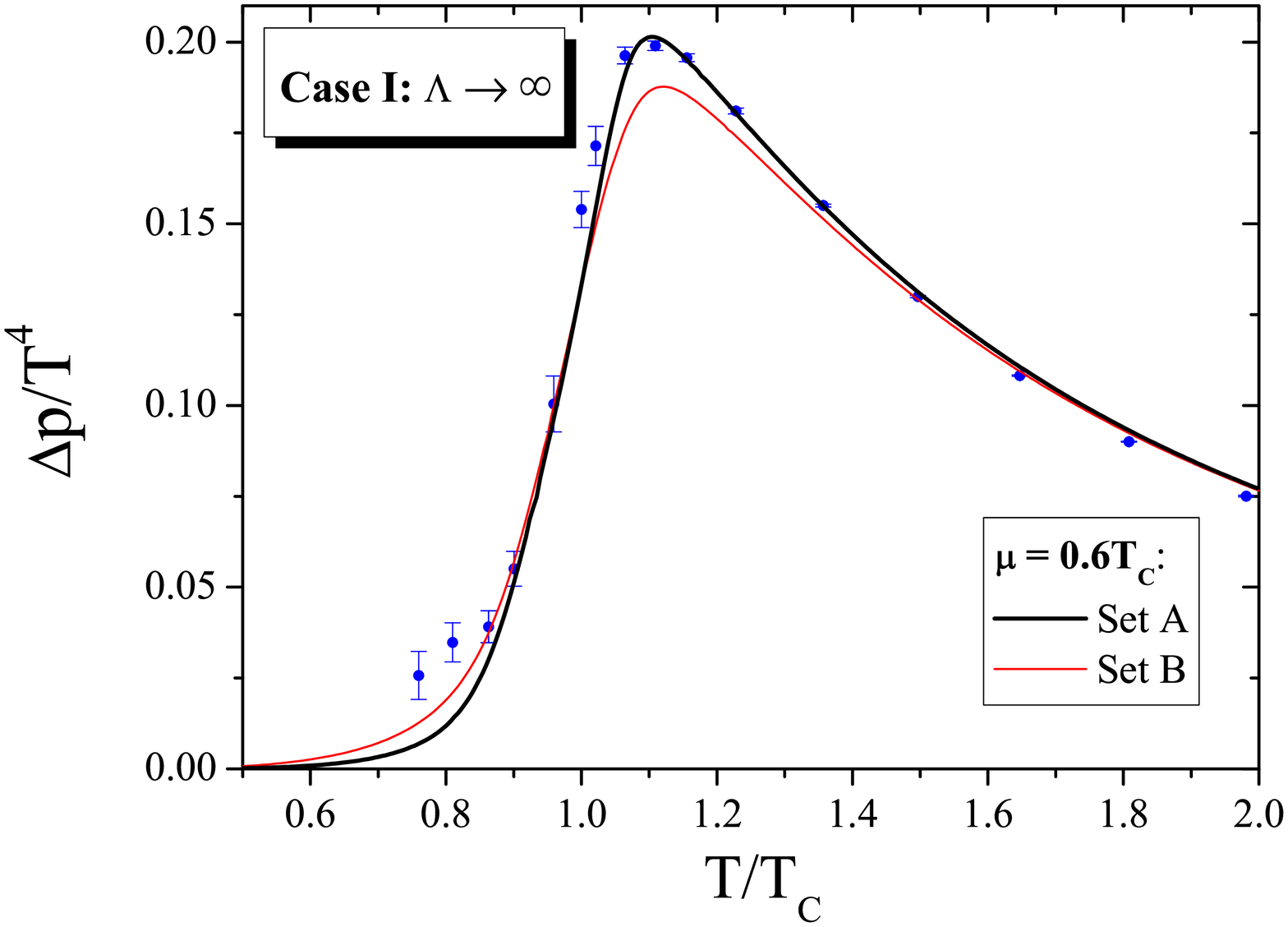,width=8.5cm,height=7.cm} &
    \hspace{-0.75cm}\epsfig{file=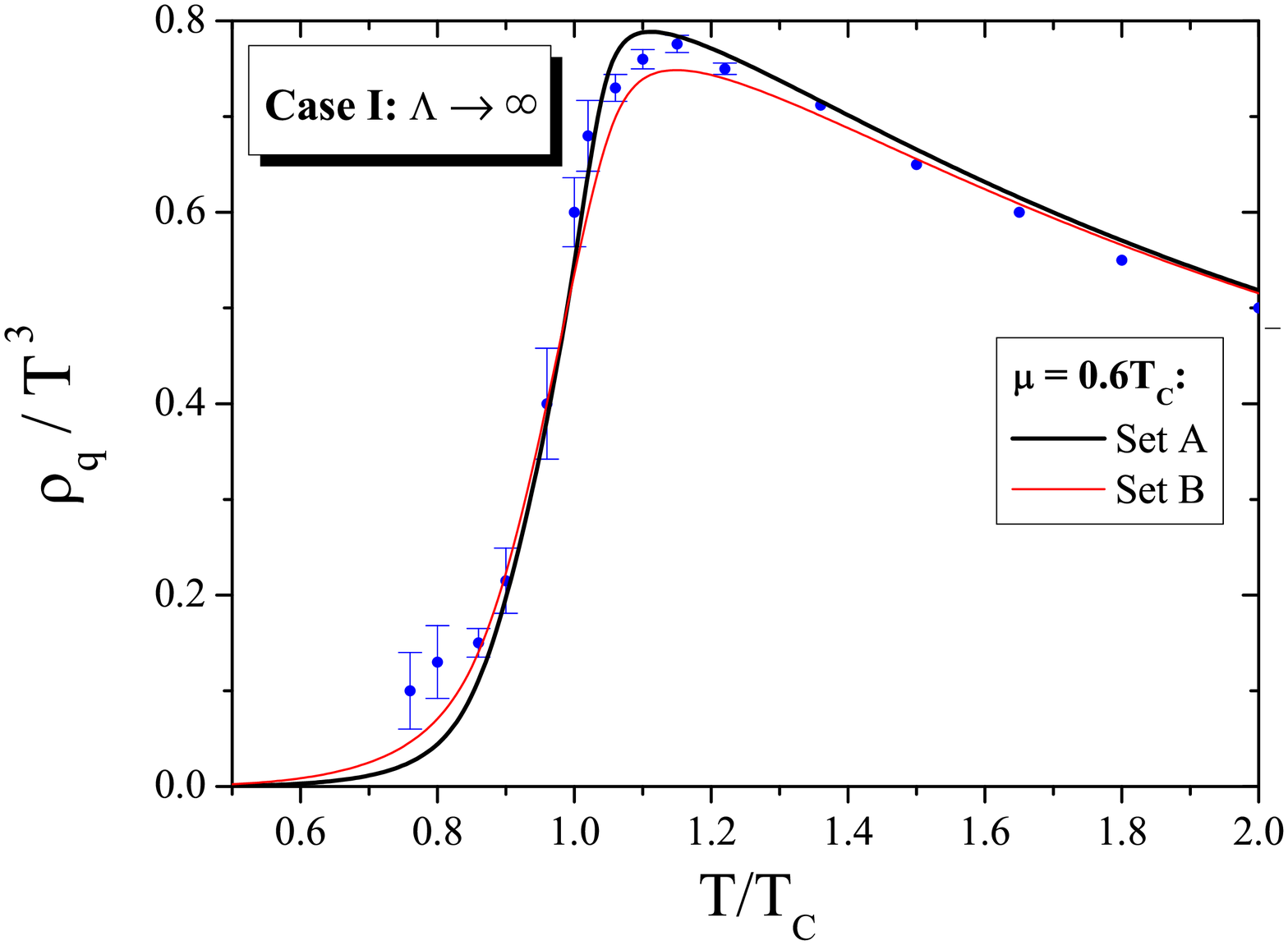,width=8.5cm,height=7.cm} \\
   \end{tabular}
\end{center}
\vspace{-0.5cm} \caption{Comparison of the scaled pressure difference (left panel) and
the scaled quark number density (right panel), as a function of temperature at finite
chemical potential for sets A and B, with the lattice data taken from Ref. \cite{lattice2}.}
\label{Fig:2}
\end{figure}

The advantage of our phenomenological model is the possibility to provide equations of
state at nonzero chemical potential too. So, we can also test its ability to reproduce
recent progress in lattice QCD with small non vanishing chemical potential.

In order to do this, we plot in Fig. 2 the scaled pressure and the quark number density
for  $\mu=0.6\,T_c$  (it was done in \cite{Ratti:2005PRD} for set B).  The parameters are
the same as in Fig. 1 and only case I for the cutoff procedure is considered. The
agreement with the lattice data is fairly good, showing that the general pattern of these
quantities and the behavior at large $T$ is reproduced. The more significative deviations
are observed in the scaled baryon number density at intermediate temperatures as is
evident from Fig. 2 (right side). However, there are some advantage with respect to the
employed parametrization A, where an improvement of the results is observed as shown in
the left panel of Fig. 2.

In conclusion, by introducing finite chemical potential we are able to compare the
obtained results with lattice data and check the validity of both sets of parameters.
From Fig. \ref{Fig:2} we conclude that, for case I, both sets of parameters are in good
agreement with lattice results, in particular, to $\Delta p/T^4$ and $\rho_q/T^3$ at
$\mu=0.6\,T_c$.


\section{Phase structure}

\begin{figure}[t]
    \begin{center}
    \begin{tabular}{cc}
        \hspace{-0.25cm}\epsfig{file=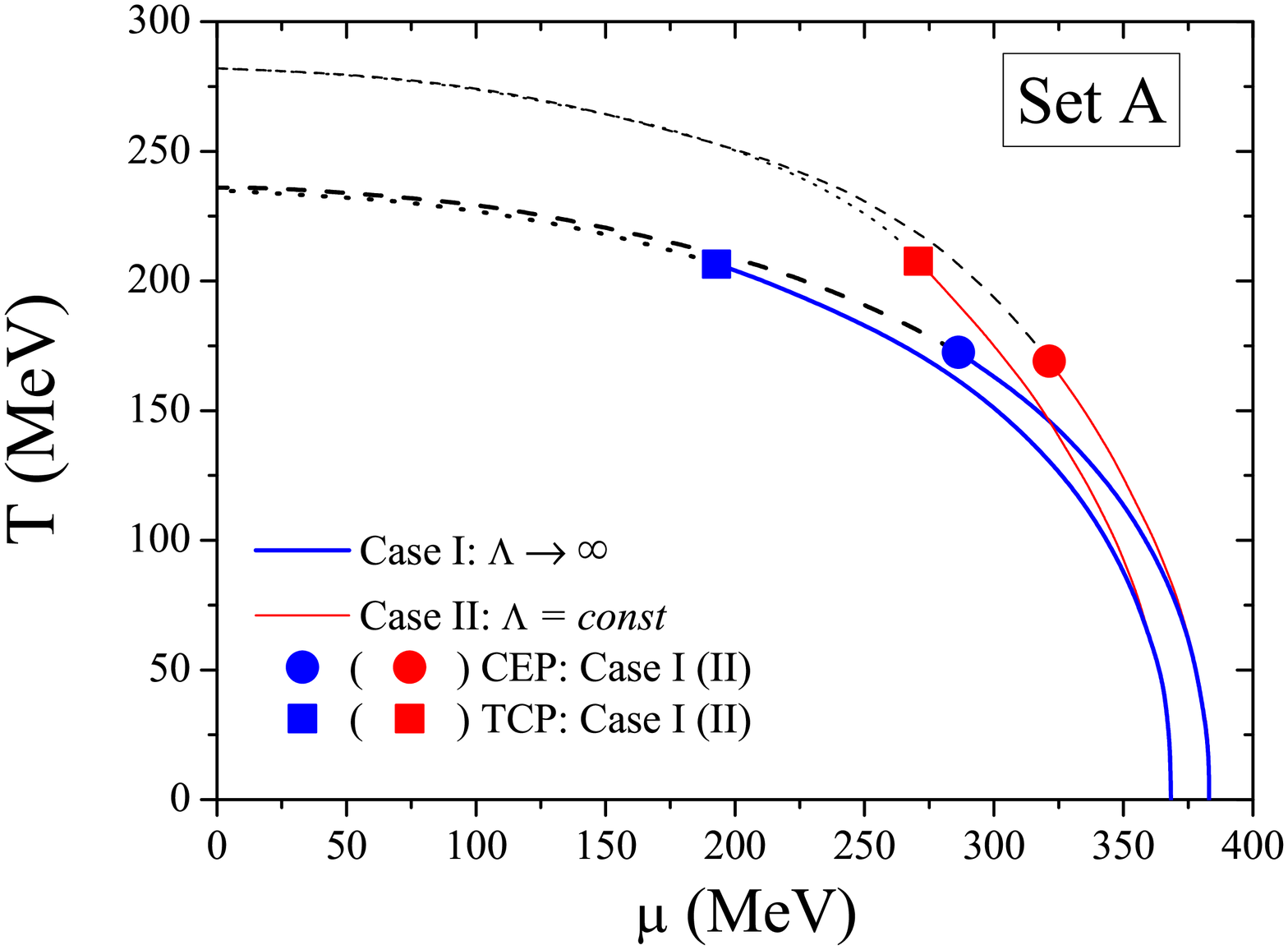,width=8.5cm,height=7.cm} &
        \hspace{-0.75cm}\epsfig{file=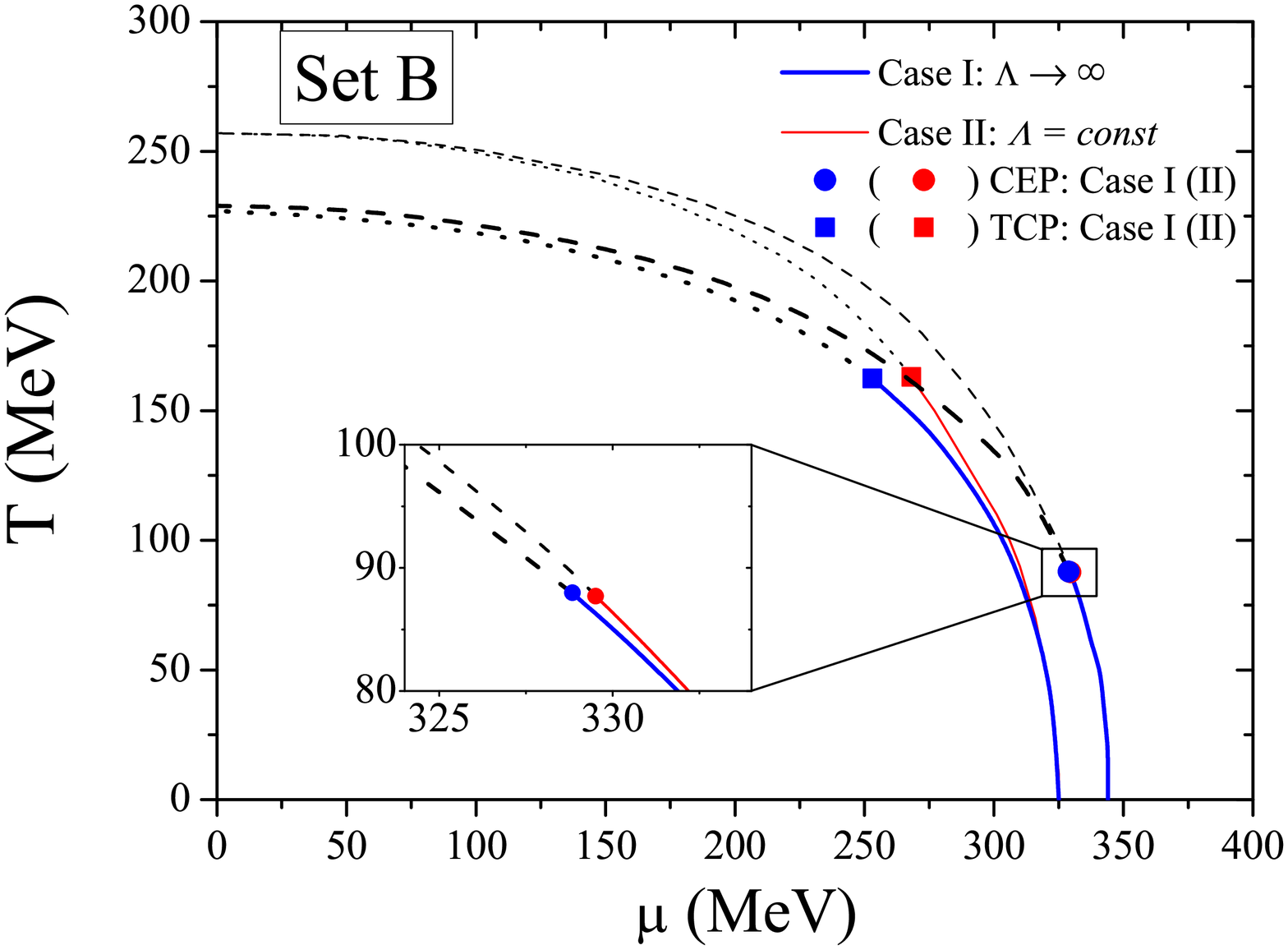,width=8.5cm,height=7.cm} \\
    \end{tabular}
    \end{center}
\vspace{-0.5cm} \caption{Phase diagrams  for cases I and II  in the chiral limit (boxes)
or not (circles) using the parameters  set A (left panel) and B (right panel). The solid
part of the curves denotes a first order transition, the dashed part to the second order
transition and the dotted line the crossover transition. } \label{Fig:3}
\end{figure}

The phase diagram for both sets of parameters (see Fig. \ref{Fig:3}) is determined by the
behavior of the orderlike parameters $\left\langle \bar q\,q\right\rangle$, 
$\Phi$ and $\bar \Phi$ together with the grand canonical potential as a function 
of temperature and chemical potential.
To draw the phase diagram we will use $T_0=270$ MeV as given by pure gauge lattice
calculations, a choice that ensures the very important physical
outcomes of lattice calculations that chiral and deconfinement
transitions coincide in the PNJL model..

We start our analysis in the limit $T=0$, where the first order phase transition occurs
at the same chemical potential for both cases I and II: $\mu = 383\,(344)$ MeV for set A
(B). This is due to the fact that at $T=0$ all the integrals are regularized with the
cutoff $\Lambda$. As a matter of fact, in this limit the Fermi functions in the gap
equations become a step function of the form $\theta(\Lambda-p_F)$ where $p_F$ is the
hadronic matter Fermi momentum. So, the integration occurs between $p_F$ and $\Lambda$
for both cases.

From Fig. 3 we also see that, at $\mu=0$, the crossover takes place for set A at
$T=235\,(272)$ MeV for case I (II), while for set B the crossover takes place at
$T=229\,(256)$ MeV for case I (II).

%
\begin{table}[t]
    \begin{center}

        \begin{tabular}{||c|c||c|c||c|c||}
        \hhline{|t:==:t:==:t:==|}
        Parameter & Regularization  & $T^{CEP}$     & $\mu^{CEP}$   & $T^{TCP}$ [MeV] & $\mu^{TCP}$ [MeV]\\
        set       & procedure       & [MeV]         & [MeV]         &(chiral limit)   & (chiral limit) \\
        \hline \hline
        \bf {Set A}   & \it{Case I: $\Lambda \rightarrow \infty$} & 172.48  & 286.35 & 206.50 & 192.87 \\
        \hline
                      & \it{Case II: $\Lambda \,=\,const$}          & 169.11  & 321.32 & 207.66 & 270.80 \\
        \hline \hline
        \bf {Set B}   & \it{Case I: $\Lambda \rightarrow \infty$} & 87.99   & 328.84 & 162.39 & 253.06 \\
        \hline
                      & \it{Case II: $\Lambda \,=\,const$}          & 87.71   & 329.51 & 163.06 & 268.05 \\
        \hhline{|b:==:b:==:b:==|}
        \end{tabular}
    \caption{\label{table:CEPTCP}
    Location of the CEP and the TCP  at the $(T,\mu)$-plane for both sets of parameters and regularization
    procedures.}
    \end{center}
    
\end{table}

At nonzero chemical potential, as the temperature increases, it is well know that the
first order transition persists up to the CEP. At the CEP the chiral transition becomes
of second order.  For temperatures above the CEP a smooth crossover takes place.

These general characteristics are  qualitatively similar for both cases, in the two sets
of parameters, as it was expected. The relevant point is   the distance between the
CEP's,  in  cases I and II, is bigger for set A than for set B. This is due to fact that
the CEP's for set A are at a higher temperature than for set B. As the temperature
increases, the high momentum quarks, that are taken into account in case I
($\Lambda\rightarrow \infty$), are more and more relevant  leading to a visible splitting
of the lines of first order phase transition and, consequently, to other location of the
CEP in the phase diagram. For set B this splitting is smaller, once  lower critical
temperatures are observed.
In the chiral limit ($m=0$ and $m_\pi=0$), the transition is of second order at $\mu=0$
and, as $\mu$ increases, the line of  second order phase transition will end in a first
order line at the TCP.
The location of the tricritical points are also included in Table \ref{table:CEPTCP}.
Nevertheless, in this case, the TCP is located at higher temperature  than the CEP (see
Fig. \ref{Fig:3}, right panel), and we already see a bigger shift between them.

\section{Nernst principle and isentropic trajectories}

The isentropic lines play a very important role in the understanding of thermodynamic
properties of matter created in relativistic heavy ion collisions. Most of the studies on
this topic   have been done on lattice calculations for two flavor QCD at finite $\mu$
\cite{Ejiri:2006PRD} but there are also studies using different type of models
\cite{Scavenius,Nonaka:2005PRC,Kahara:2008}. Some model calculations predict that  in a
region around the CEP the properties of matter are only slowly modified as the collision
energy is changed, as a consequence of the attractor character of the CEP
\cite{Stephanov:1998PRL}.

\begin{figure}[t]
\begin{center}
  \begin{tabular}{cc}
    \hspace{-0.25cm}\epsfig{file=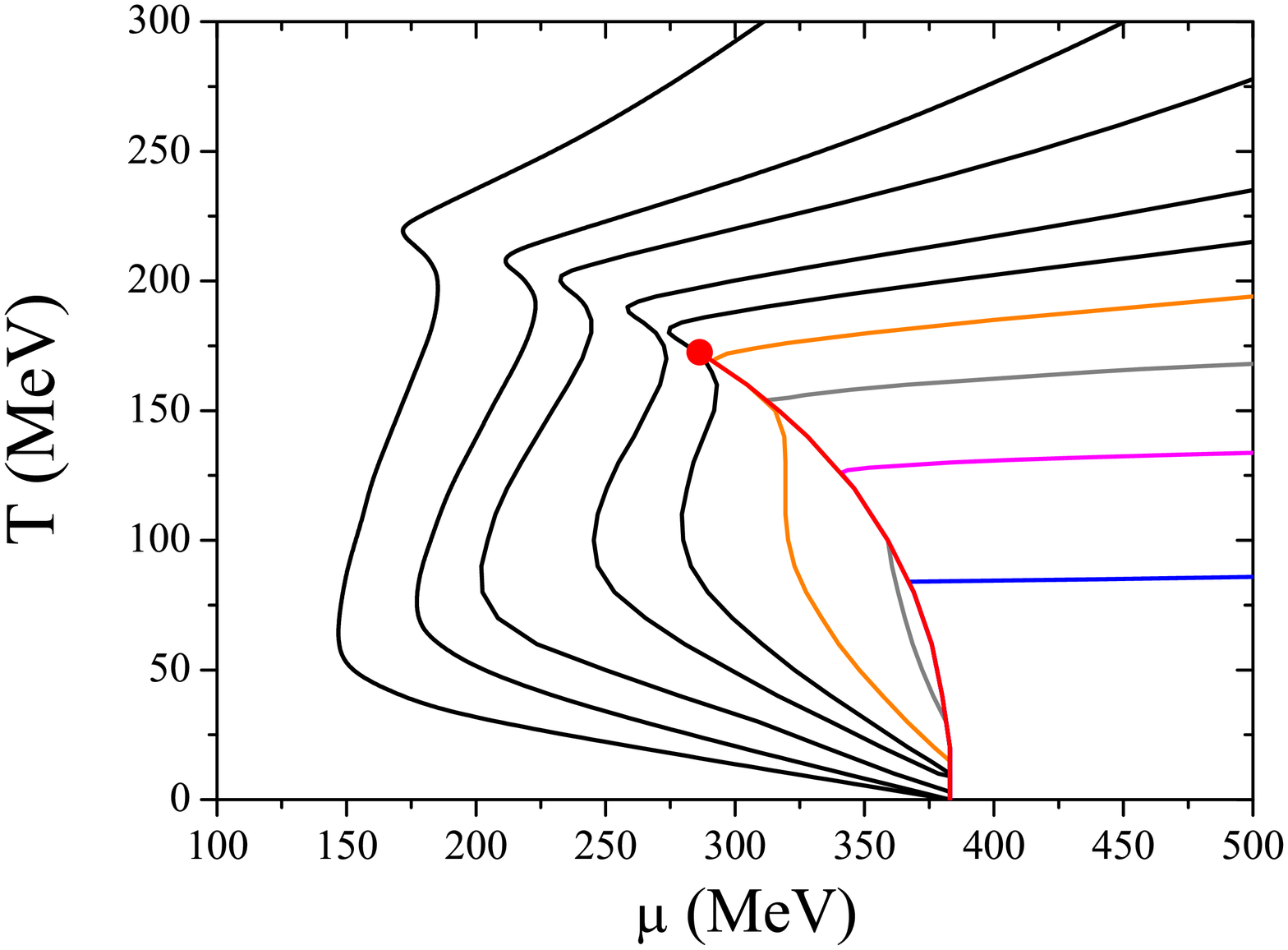,width=8.5cm,height=7.cm} &
    \hspace{-0.75cm}\epsfig{file=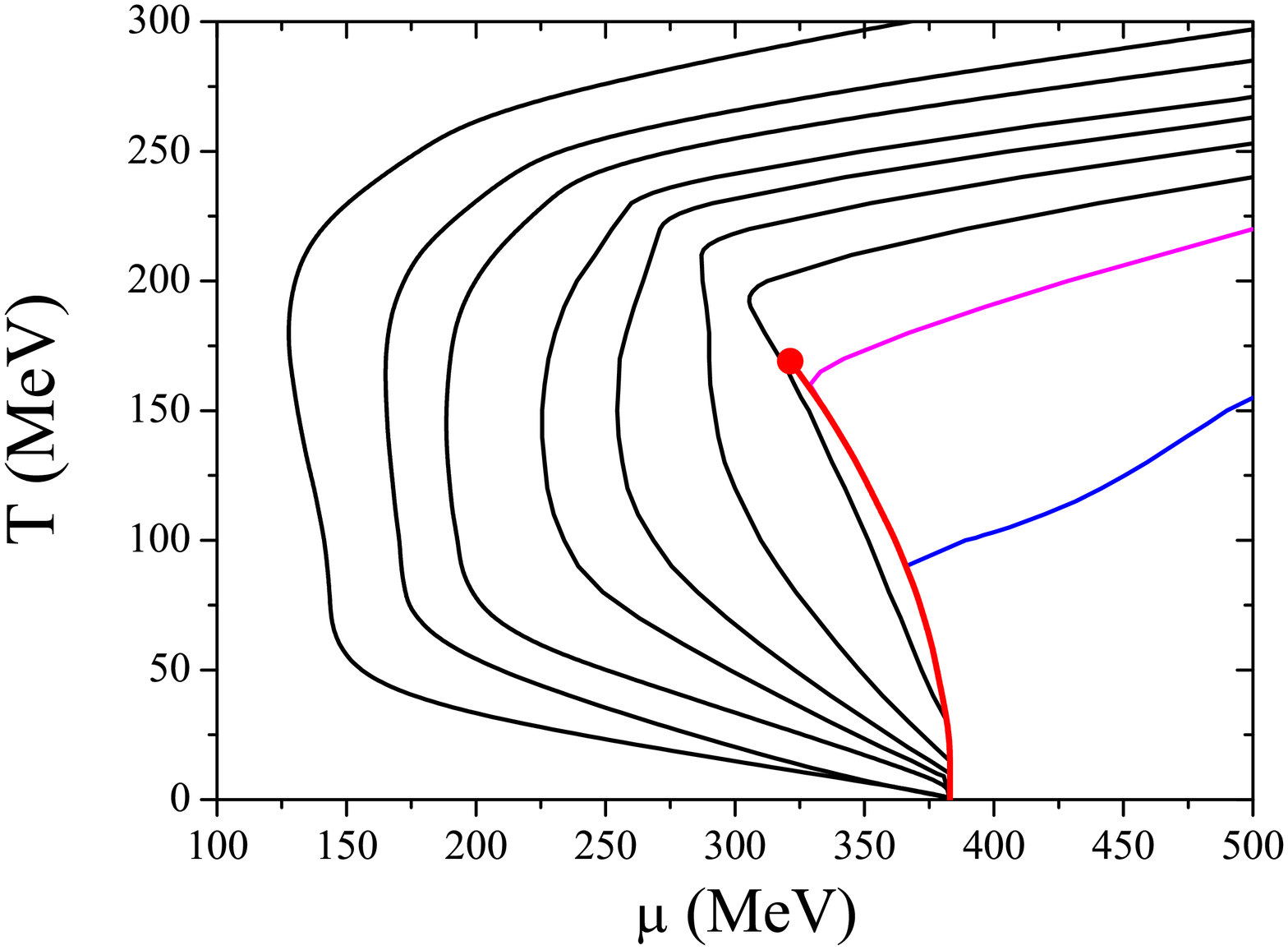,width=8.5cm,height=7.cm} \\
  \end{tabular}
\end{center}
\vspace{-0.5cm} \caption{Isentropic trajectories  in the $(T,\mu)$-plane for case I (left
panel) and case II (right panel) using the parameter set A. The following values of the
entropy per baryon number have been  considered:
$s/\rho_q=1,\,2,\,3,\,4,\,5,\,6,\,8,\,10,\,15$ (anticlockwise direction). } 
\label{Fig:4}
\end{figure}

Our numerical results  for the isentropic lines in the $(T,\mu)$-plane are shown in Fig.
\ref{Fig:4}, where we have used  set A of parameters and both regularization procedures.

We start the discussion  by analyzing the behavior of the isentropic lines in the limit
$T\rightarrow 0$.  We point out that, as already referred by other authors
\cite{Scavenius}, in this limit:
\begin{itemize}
  \item [(i)] $s \rightarrow 0$, according to the third law of thermodynamics; and
  \item [(ii)] for $s/\rho_q\,=\,const$, we have to insure that also $\rho_q \rightarrow 0$.
\end{itemize}

However, the satisfaction of the condition (ii) is only provided when $\mu\leq M_{vac}$.
In spite of the general use of set B in the literature of the PNJL model, only set A
satisfies  this ansatz.
We remember (Sec. 2.3)  that with  set A we are, at $T=0$, in the presence of droplets
(states in mechanical equilibrium with the vacuum state at $P=0$).

Consequently, even without reheating in the mixed phase as
verified in the ``zigzag'' shape of
\cite{Ejiri:2006PRD,Nonaka:2005PRC,Kahara:2008,Subramanian}, all isentropic trajectories
directly terminate at the first order transition line at $T=0$. So, for set A it is
verified that $s \rightarrow 0$ and $\rho_q \rightarrow 0$  in the limit $T\rightarrow
0$,  as it should be.

In conclusion, our convenient choice of the model parameters allows a first order phase
transition that is stronger than in other treatments of the NJL (PNJL) model. This choice
is crucial to obtain important results:  the criterium of stability of the quark droplets
\cite{Buballa:2004PR,Costa:2003PRC} is fulfilled, and, in addition,  simple thermodynamic
expectations in the limit $T\rightarrow 0$ are verified.

At $T\neq0$, in the first order line, the behavior we find is somewhat different from
those claimed by other authors \cite{Nonaka:2005PRC,Stephanov:1999PRD} where a phenomena
of focusing of trajectories towards the CEP is observed. For case I (see Fig.
\ref{Fig:4}, left panel) we see that the isentropic lines with $s/\rho_q=1,...,4$ come
from the region of symmetry partially restored and attain directly the phase transition,
going along with the phase transition as $T$ decreases until it reaches $T=0$. The same
behavior is found for case II when $s/\rho_q=1,2$ (see Fig. \ref{Fig:4}, right panel).
For case II, we also observe, in a small range of $s/\rho_q$ around $3$, a tendency to
convergence of these isentropic lines towards the CEP. These lines come from the region
of symmetry partially restored in the direction of the crossover line. For smaller values
of $s/\rho_q$, the isentropic lines turn about the CEP and then attain the first order
transition line. For larger values of $s/\rho_q$ the isentropic trajectories approach the
CEP by the region where the chiral symmetry is still broken, and also attain the first
order transition line after bending toward the critical point.
As already pointed out in \cite{Scavenius}, this is a natural result in these type of
quark models with no change in the number of degrees of freedom of the system in the two
phases. As the temperature decreases a first order phase transition occurs, the latent
heat increases and  the formation of the mixed phase is thermodynamically favored.

In the crossover region, for both cases, the behavior of the isentropic lines is
qualitatively similar to the one obtained in lattice calculations \cite{Ejiri:2006PRD} or
in some models \cite{Nonaka:2005PRC,Fukushima,Nakano}. 
On the other hand, the isentropic trajectories in the phase diagram indicate that the 
slope of the trajectories goes to large values for large $T$.
We can also conclude that, in the PNJL model, the entropy and the baryon number density
are very sensitive to the regularization procedure used \cite{Klevansky,Costa:2008PRD2},
and this effect is also relevant for the present situation.


\section{Susceptibilities and critical exponents}

The grand canonical potential (or the pressure) contains the relevant information on
thermodynamic bulk properties of a medium. Susceptibilities, being second order
derivatives of the pressure in both chemical potential and temperature directions, are
related to fluctuations that are supposed to represent signatures of phase transitions of
strongly interacting matter. In particular,  the quark number susceptibilities play a
role in the calculation of event-by-event fluctuations of conserved quantities such as
net baryon number. Across the quark hadron phase transition they are expected to become
large, what can be interpreted as  an indication for a critical behavior. We also
remember the important role of the second derivative of the pressure for second order
points like the CEP.

In previous works \cite{Costa:2008PRD3,Costa:2008PRD1}, we have studied the CEP within
the restrictions imposed by the regularization in case II. It is important to investigate
if the type of regularization plays a significant role in the critical properties of
physical observables, such as the baryon number susceptibility and the specific heat, and
respective critical exponents, in the vicinity of the CEP. The relevance of these
physical observables is due to the size of the critical region around the CEP which can
be found by calculating the baryon number susceptibility, the specific heat and their
critical behaviors. The size of this critical region is important for future searches of
the CEP in heavy ion-collisions \cite{{Nonaka:2005PRC}}.

\begin{figure}[t]
\begin{center}
    \epsfig{file=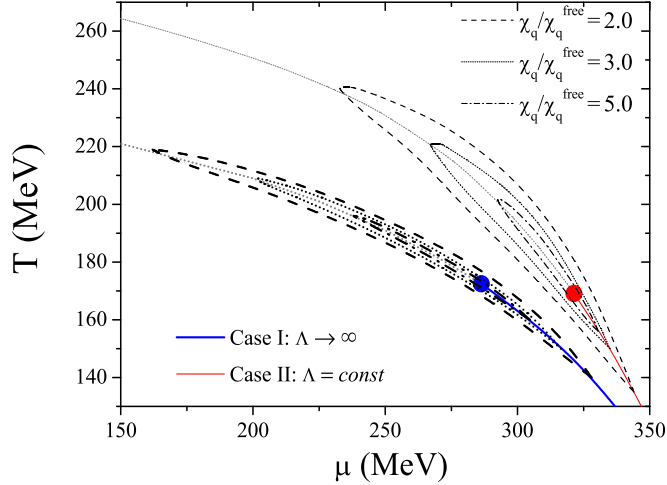,width=10.5cm,height=7.5cm}
\end{center}
\vspace{-0.5cm} \caption{Phase diagram in the PNJL (case I and case II) for the parameter
set A. The size of the critical region around the CEP is plotted for
$\chi_q/\chi_q^{free}=2,3,5$.} \label{Fig:5}
\end{figure}

In our calculations, we will use only the set A of parameters due to the advantages of
this set as explained in the previous sections. In Fig. \ref{Fig:5} we plot the phase
diagram in a region around the CEP for both cases.

The way to estimate the critical region around the CEP is to calculate the dimensionless
ratio $\chi_q/\chi_q^{free}$, where $\chi_q^{free}$ is obtained taking the chiral limit
$m = 0$. Figure \ref{Fig:5} shows a contour plot for three fixed ratios
($\chi_q/\chi_q^{free}=2.0,\,3.0,\,5.0$) in the phase diagram around the CEP. We notice
an elongation, in the direction parallel to the first order transition line, of the
region where $\chi_q$ is enhanced, indicating that the critical region is heavily
stretched in that direction.

The elongation of the critical region in the $(T,\mu)$-plane, along the line of the phase
transition (or the crossover), is larger in case I (see for example $\chi_q/\chi_q^{free}
= 2.0$ in Fig. \ref{Fig:5}). It means that the divergence of the correlation length at
the CEP affects the phase diagram quite far from the CEP, particularly in case I, and a
careful analysis including effects beyond the mean-field needs to be done
\cite{Rossner:2007ik}.

We remember  that  one of the main effects of the Polyakov loop is to shorten the
temperature range where the crossover occurs \cite{Hansen:2007PRD}. On the other hand,
this behavior is boosted by using $\Lambda \rightarrow \infty$: at $\mu = 0$ the
crossover occurs between about 180 MeV and 270 MeV, for case I, and between about 210 MeV
and 325 MeV, for case II. The combination of both effects results in higher baryonic
susceptibilities even far from the CEP.
This effect of the Polyakov loop is driven by the fact that the one- and two-quark
Boltzmann factors are controlled by a factor proportional to $\Phi$: at small temperature
$\Phi \simeq 0$ results in a suppression of these contributions (see Eq. \ref{omega})
leading to a partial restoration of the color symmetry. Indeed, the fact that only the
$3-$quarks Boltzmann factors $e^{3\beta E_p}$ contribute to the thermodynamical potential
at low temperature, may be interpreted as the production of a thermal bath containing
only colorless 3-quarks contributions.  When the temperature increases, $\Phi$ goes
quickly to 1 (this is faster in case I due to the higher momentum quarks present in the
system) resulting in a (partial) restoration of the chiral symmetry occurring in a
shorter temperature range.
The crossover taking place in a smaller $T$ range can be interpreted as a crossover
transition closest to a second order one, a feature that is more clear in case I than in
case II. This ``faster'' crossover may explain the elongation of the critical region in
case I, compared to case II, giving raise to a greater correlation length even far from
the CEP.

Now, we will investigate the behavior of $\chi_q$ and $C$ in the vicinity of the CEP and
their critical exponents, for both cases. The calculated critical exponents at the CEP
and the TCP, together with the universality/mean-field predictions, are presented in
Table \ref{table:critexpo} and will be discussed in the sequel.

\begin{table}[t]\label{Tab:1}
\begin {center}
\small{
\begin{tabular}{ccccc}
    \hline
    {Quantity} & {critical exponents/path} & {Case I} & {Case II} & {Universality} \\
    \hline \hline
  & $\epsilon\,/\,\,{\rm from \,left \,( {\textcolor{red}\Rightarrow})}$ & { {$0.66 \pm 0.01$}}
  & {$0.66 \pm 0.01$} & {$2/3$} \\
  {$\chi_q$} & {{$\epsilon^\prime\,/\,$ from right\,$({\textcolor{red}\Leftarrow})$}} &
  {$0.69 \pm 0.01$} & {$0.69 \pm 0.02$} & {$2/3$} \\
  &  $\gamma_q$
$\,/$ from left $({\textcolor{blue}\rightarrow})$ &
     $0.53 \pm 0.02$ & $0.51 \pm 0.01$ &  {$1/2$}  \\
    \hline
    & $\alpha\,/$ from below

          $ {({\textcolor{red}\Uparrow)}}$

    & {$
        \begin{array}{c}
          0.64\pm 0.01 \\
         \alpha_1= 0.58 \pm 0.01
        \end{array}$}
    & {$
        \begin{array}{c}
          0.62\pm 0.02 \\
          \alpha_1=0.52\pm 0.01
        \end{array}$}
    & {$
        \begin{array}{c}
        2/3 \\
        $---$
        \end{array}$} \\
  {$C$}
  & $\alpha^\prime/$ from above

$({\textcolor{red}\Downarrow})$

    & {$0.68 \pm 0.01$} & {$0.68 \pm 0.01$} & {$2/3$} \\
    & $\alpha$  $\,/$ from below
        $({\textcolor{blue}\uparrow})$

    & {$0.50 \pm 0.01$} & {$0.47 \pm 0.02$} & {$1/2$}  \\
\hline
\end{tabular}
\caption{
\label{table:critexpo}
The arrow ${\textcolor{red}\Rightarrow}({\textcolor{blue}
  \uparrow})$ indicates the path in the $\mu\,(T)-$ direction to the 
  {\textcolor{red}{CEP}}/{\textcolor{blue}{TCP}}
   for ${\mu<\mu^{CEP}}$
$({T<T^{TCP}}$).
}
}
\end{center}
\end{table}

\begin{figure}[t]
    \begin{center}
    \begin{tabular}{cc}
        \hspace{-0.25cm}\epsfig{file=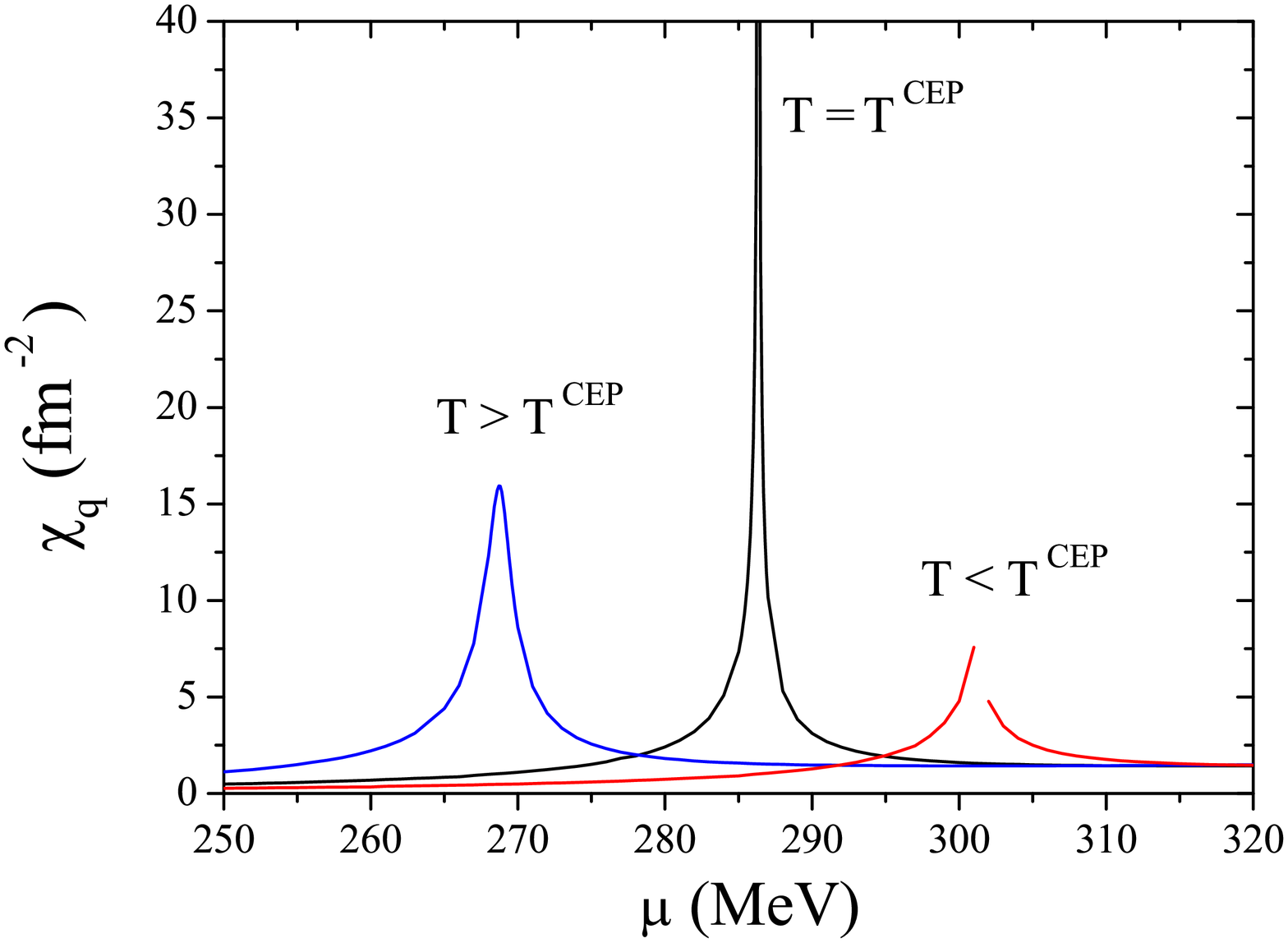,width=8.5cm,height=7.5cm} &
        \hspace{-0.75cm}\epsfig{file=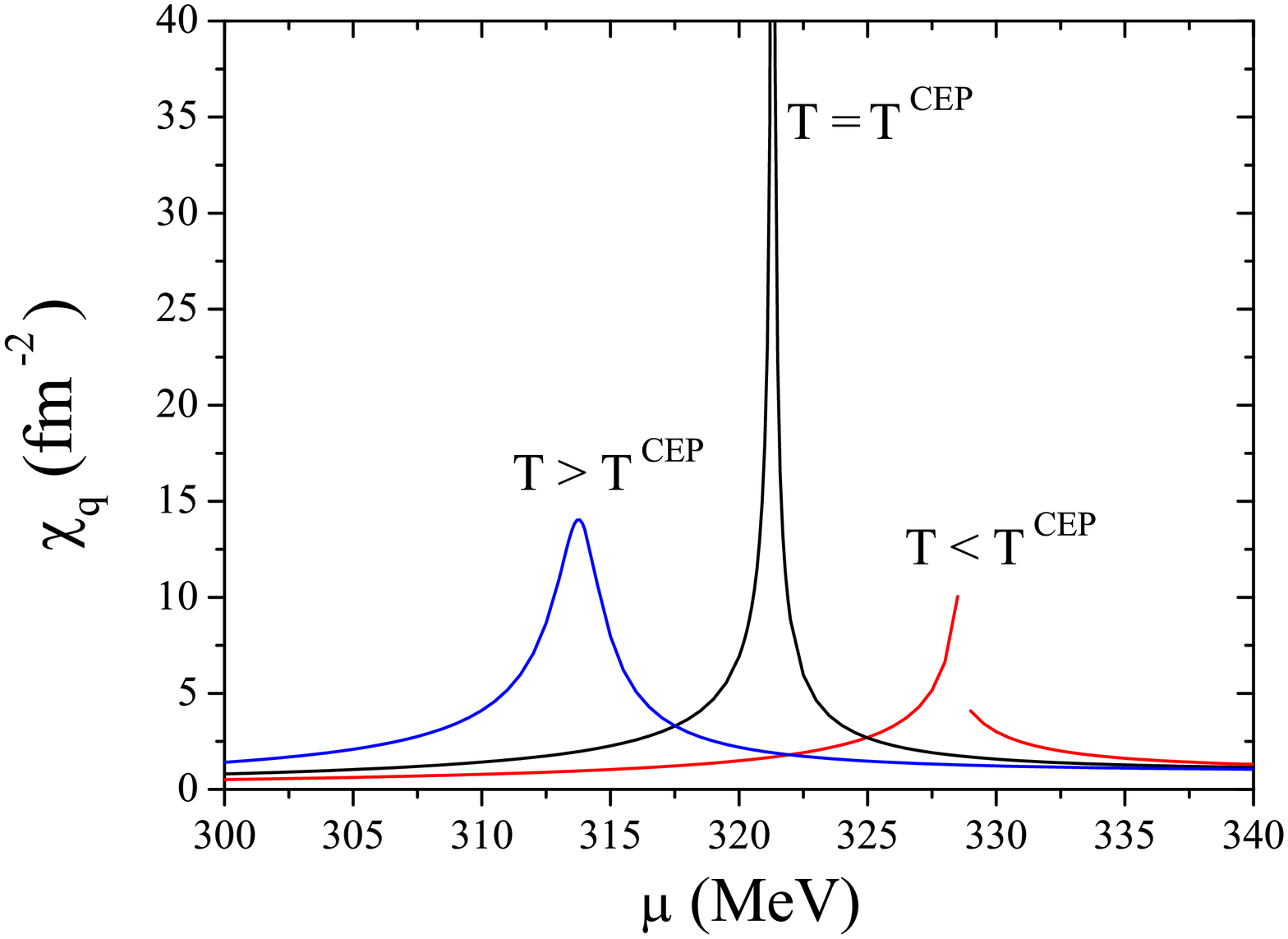,width=8.5cm,height=7.5cm} \\
    \end{tabular}
    \end{center}
\vspace{-0.5cm} \caption{Left panel: Baryon number susceptibility for case I as function
of $\mu$ for different temperatures around the CEP: $T^{CEP}=172.48$ MeV and
$T=T^{CEP}\pm10$ MeV. Right panel: Baryon number susceptibility for case II as function
of $\mu$ for different temperatures around the CEP: $T^{CEP}=169.11$ MeV and
$T=T^{CEP}\pm10$ MeV.} \label{Fig:6}
\end{figure}

In the left panel of Fig. \ref{Fig:6} the baryon number susceptibility is
plotted as a function of $\mu$ for three different temperatures around the CEP in the
context of case I. The behavior is very similar in both cases. For  $T<T^{CEP}$, the
phase transition is first order and  $\chi_q$ has a discontinuity; for $T = T^{CEP}$, the
slope of the baryon number density tends to infinity at $\mu=\mu^{CEP}$ and  $\chi_q$
diverges; for $T>T^{CEP}$, the discontinuity of $\chi_q$ disappears at the transition
line. A similar behavior of $\chi_q$ is found for case II, as we can see from the right
panel of Fig. \ref{Fig:6}.

\begin{figure}[t]
\begin{center}
  \begin{tabular}{cc}
    \hspace{-0.25cm}\epsfig{file=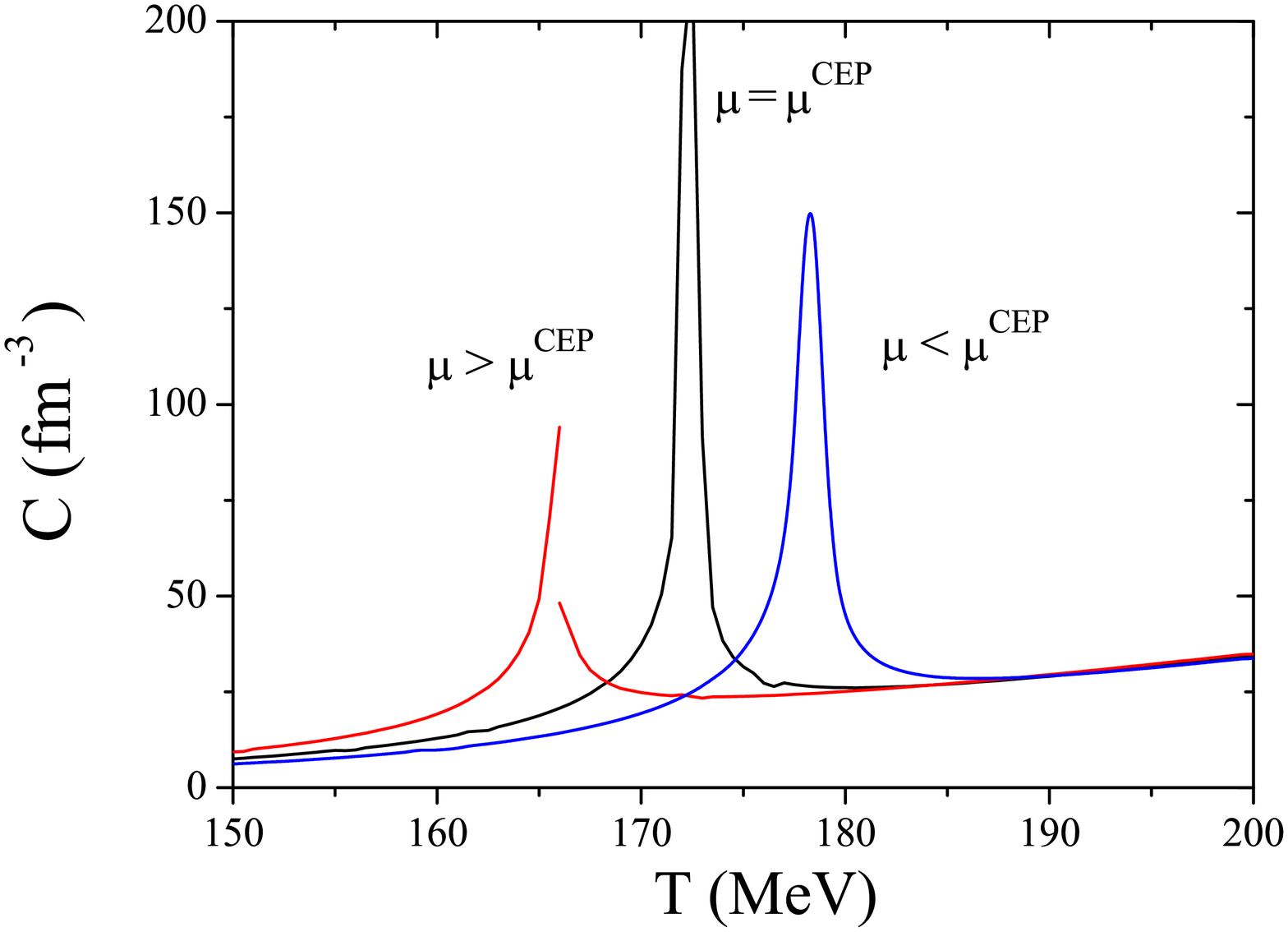,width=8.5cm,height=7.5cm} &
    \hspace{-0.75cm}\epsfig{file=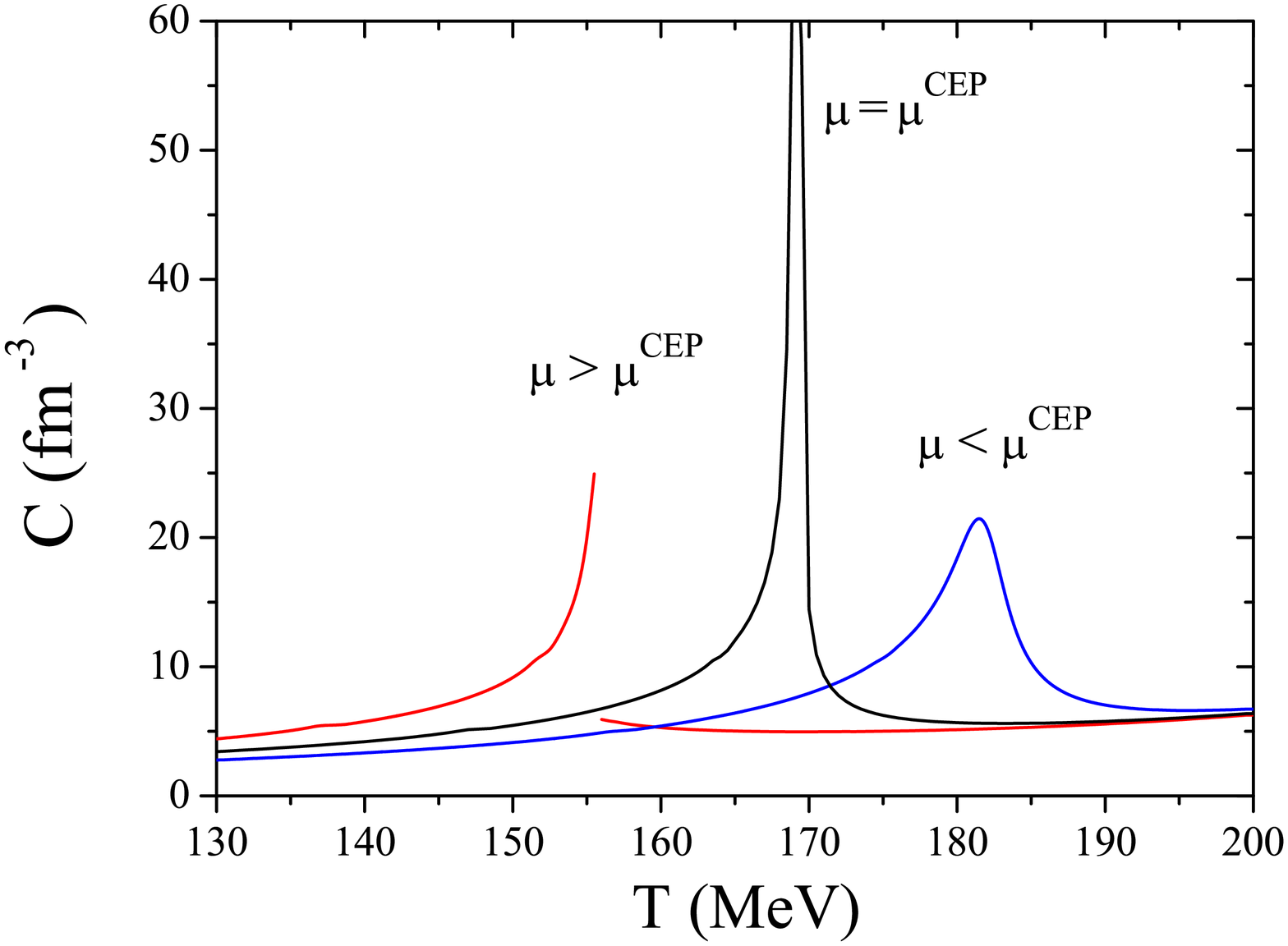,width=8.5cm,height=7.5cm} \\
  \end{tabular}
\end{center}
\vspace{-0.5cm} \caption{Left panel: Specific heat for case I as a function of $T$ for
different values of $\mu$ around the CEP: $\mu^{CEP}=286.35$ MeV and $\mu=\mu^{CEP}\pm10$
MeV. Right panel: Specific heat for case II as a function of $T$ for different values of
$\mu$ around the CEP: $\mu^{CEP}=321.32$ MeV and $\mu=\mu^{CEP}\pm10$ MeV. }
\label{Fig:7}
\end{figure}

The behavior of the specific heat for both cases, as a function of temperature for three
different chemical potentials around the CEP, is presented in  Fig. \ref{Fig:7}. The
behavior found for $C$ around the CEP is very similar to the behavior of $\chi_q$ for
both Cases as we can see from Fig. \ref{Fig:7}.

It is interesting to notice that the high momentum quarks introduced in case I ($\Lambda
\rightarrow \infty$), and that are not taken  into account in case II,  have no
significant effect on both $\epsilon$ and $\alpha$. However, we observe that the peak at the
critical points $T^{CEP}$ or $\mu^{CEP}$ is sharper in the PNJL model in case I (as it
can be expected from the analysis of the stretching of the critical region done above).

To better understand the extreme behavior of  $\chi_q$ and $C$ near the CEP, we will
determine the critical exponents (in our case $\epsilon$ and $\alpha$ are the critical
exponents of $\chi_q$ and $C$, respectively).
These critical exponents will be determined by  following two directions,
temperature-like and magnetic-field-like, in the $(T,\mu)$-plane near the CEP, because,
as pointed out in \cite{Griffiths:1970PR}, the form of the divergence depends on the
route that is chosen to approach the CEP.

Starting with the baryon number susceptibility, for both cases, if the path chosen is
asymptotically parallel to the first order transition line at the CEP, the divergence of
$\chi_q$ scales with an exponent $\gamma_q$. In the mean-field approximation it is
expected to find $\gamma_q=1$ for this path. For directions not parallel to the tangent
line, the divergence scales as $\epsilon =2/3$. These values are responsible for the
elongation of the critical region, $\chi_q$ being enhanced in the direction parallel to
the first order transition line (see Fig. \ref{Fig:5}).

To study the critical exponents for the baryon number susceptibility (Eq. \ref{chi}) we
will start with a path parallel to the $\mu$-axis in the ($T,\mu$)-plane, from lower
$\mu$ towards the CEP ($T^{CEP},\mu^{CEP}$).
Using a linear logarithmic fit
$  \ln \chi_q = -\epsilon \ln |\mu -\mu^{CEP} | + c_1$,
where the term $c_1$ is independent of $\mu$, we obtain $\epsilon = 0.66\pm 0.01$, which
is consistent with the mean-field theory prediction, $\epsilon = 2/3$. This value is also
similar to the value found in case II as we can see from Table \ref{table:critexpo} 
(see also Ref. \cite{Costa:2008PRD3}).

We also study the baryon number susceptibility from higher $\mu$ towards the critical
$\mu^{CEP}$ for both cases. The logarithmic fit used now is $\ln \chi_q = -\epsilon' \ln
|\mu -\mu^{CEP}| + c'_1$. Our result for case I shows that $\epsilon'  = 0.69\pm
0.01\approx \epsilon$. This means that the size of the region we observe is approximately
the same independently of the direction we choose for the path parallel to the
$\mu$-axis.
 Once again, this value is similar to the critical exponent for case II and
is consistent with the mean-field theoretical prediction $\epsilon = 2/3$ (see the
results presented in Table \ref{table:critexpo}).

On the other hand, in the chiral limit (where the CEP becomes a TCP), it is  found that
the critical exponent for $\chi_q$, denoted by $\gamma_q$, has the value $\gamma_q=0.53$
$\pm 0.02$ ($\gamma_q=0.51$  $\pm 0.01$), for case I (II). Again, these results are in
agreement with the mean-field value ($\gamma_q=1/2$).

Let us pay attention to the specific heat around the CEP. We can calculate the critical
exponent using a path parallel to the $T$-axis in the ($T,\mu$)-plane from lower
$T$ towards the CEP.
We  observe that, as already found in Ref. \cite{Costa:2008PRD3} for case II, the slope
of data points change for  $|T-T^{CEP}|$ around $0.3$ MeV. So, for case I (II) we obtain
the critical exponents $\alpha=0.64\pm 0.01\, (0.62\pm 0.02)$ and $\alpha_1=0.58\pm
0.01\,(0.52\pm 0.01)$. As pointed out in \cite{Hatta:2003PRD}, this change of the
exponent can be interpreted as a crossover of different universality classes, with the
CEP being affected by the TCP due to the small physical quark masses.

The value of $\alpha$ in both cases is consistent with the one suggested by universality
arguments in \cite{Hatta:2003PRD}: it is expected that $\chi_q$ and $C$ should be
essentially the same near the TCP and the CEP which implies $\alpha=\epsilon=2/3$.
In addition, if we compare $\alpha_1$ for both cases, we conclude that $\alpha_1$ in
case I is closer to $\alpha$ than in case II. It seems that the high momentum quarks
allowed in case I affect the crossover of different universality classes making
$\alpha_1$ getting closer to $\alpha$ which is already consistent with
$\alpha=\epsilon=2/3$

When the critical point is approached from above the following exponents are obtained:
$\alpha^\prime = 0.68\pm 0.01$ $(0.68\pm 0.01)$ for case I (II).

Finally, concerning the  behavior of the specific heat around the TCP, we find as shown
in Table \ref{table:critexpo}, $\alpha=0.50\pm 0.01$ for case I and $\alpha=0.47\pm 0.02$ for case II. 
These values are in agreement with the respective mean-field value ($\alpha=1/2$).


\section{Conclusions}

We have considered the  PNJL model as one of the prototype models of dynamical symmetry
breaking  of QCD (both chiral and ``color'' symmetry) and investigated the phase
structure at finite $T$ and $\mu$. Evaluating the thermodynamical potential we find the
critical curves on the $(T,\mu)$-plane. Working out of the chiral limit, a CEP which
separates the first and the crossover line is found. First and second derivatives of the
thermodynamical potential are also evaluated.

The critical phenomena related with the explicit breaking and partial restoration of
$\Z_3$ and chiral symmetry are analyzed.  Two different sets of parameters have been used
with the emphasis on the parameter choice which is compatible with the formation of
stable droplets at zero temperature.
The effects of two regularization procedures at finite temperature, one that allows high
momentum quark states to be present (I) and the other not (II), have also been discussed.

We have found that the  presence of the Polyakov loop provides a substantial enhancement
of the  critical temperature, bringing it to a better agreement with the most recent
results of lattice calculations. Another interesting effect of the coupling of quarks to
the Polyakov loop is that the phase transition becomes steeper, showing a sharper peak in
the baryon number  susceptibility and the specific heat. This effect is reinforced when
regularization I is used.

The observation of differential observables, like the entropy and the heat capacity and
its temperature and density behavior,   serve as important probes that, together with
lattice data, are important for the phenomenology of heavy-ion collisions and cosmology.
In both cases A and B, as well as in both cutoff procedures, the gross structure of the
phase diagram expected for QCD is obtained.  The location of the CEP depends on the model
parameters  and, in the chiral limit, a TCP is found according to universality arguments.

Two important points of our model calculation concern the  choice of the model parameters
and  the regularization procedure at finite temperature as already referred.
We conclude that the choice of the model parameters has  important consequences in order
to obtain the correct asymptotic low temperature behavior. In the zero temperature limit,
the chemical potential approaches a finite value that must satisfy to the condition
$\mu_c<M_{vac}$. Only the set of parameters  A  insures this condition that allows us to
obtain both $s=0$ and $\rho_q=0$.
The regularization procedure is important for obtaining agreement with the asymptotic
behavior above $T_c$.

Observables like $\chi$ and $C$,  which are obtained as derivatives of the
thermodynamical potential with respect to $\mu$ and $T$ respectively, allow us to explore
the effects of the Polyakov loop on the thermodynamical properties.
The successful comparison with lattice results shows that the model calculation provides
a convenient tool to obtain information for  systems from zero to nonzero chemical
potential which is of particular importance for the knowledge of the equation of state of
hot matter, relevant for the upcoming LHC experiments at CERN,  and for dense matter,
relevant for the CBM one at FAIR.

\begin{acknowledgments}
Work supported by grant SFRH/BPD/23252/2005 (P. Costa), Centro de F\'{\i}sica
Computacional, FCT under projects POCI/FP/81936/2007 (H. Hansen) and CERN/FP/83644/2008.
\end{acknowledgments}
\vspace{0.5cm}



\begin{thebibliography}{99}

\bibitem{HalaszJSSV}
        M. A. Halasz, A. D. Jackson, R. E. Shrock, M. A. Stephanov, and J. J. M. Verbaarschot,
        Phys. Rev. D {\bf 58},096007 (1998).

\bibitem{Shuryak2006}
        J. Liao, and E. Shuryak,
        Nucl. Phys. {\bf A775}, 224 (2006).

\bibitem{Barducci:1994PRD}
        A. Barducci {\it et al.},
        Phys. Lett. B {\bf 231}, 463 (1989);
        Phys. Rev. D {\bf 41}, 1610 (1990);
        A. Barducci {\it et al.},
        Phys. Rev. D {\bf 49}, 426 (1994).

\bibitem{Stephanov:1996PRL}
        M. A. Stephanov,
        Phys. Rev. Lett. {\bf 76}, 4472 (1996).

\bibitem{Schwarz:PRC1999}
        T. M. Schwarz, S. P. Klevansky, and G. Papp,
        Phys. Rev. C {\bf 60}, 055205 (1999).

\bibitem{Buballa:2003PLB}
        M. Frank, M. Buballa, and M. Oertel,
        Phys. Lett. B {\bf 562}, 221 (2003).

\bibitem{Costa:2007PLB}
        P. Costa, C. A de Sousa, M. C. Ruivo, and Yu. L. Kalinovsky,
        Phys. Letts. B {\bf 647}, 431 (2007);
        P. Costa, C. A. de Sousa, and M. C. Ruivo,
        Phys. Rev. D {\bf 77}, 096001 (2008).

\bibitem{Asakawa}  
				M. Asakawa, 
				J. Phys. G: Nucl. Part. Phys. {\bf 36}, 064042 (2009).

\bibitem{Stephanov:1998PRL}
        M. Stephanov, K. Rajagopal, and E. Shuryak,
        Phys. Rev. Lett. {\bf 81}, 4816 (1998).

\bibitem{Hatta:2003PRD}
        Y. Hatta and T. Ikeda,
        Phys. Rev. D {\bf 67}, 014028 (2003);
        B.-J. Schaefer and J. Wambach,
        Phys. Rev. D {\bf 75}, 085015 (2007).

\bibitem{Karsch}
				Z. Fodor and S. D. Katz, 
        J. High Energy Phys. {\bf 04}, 050 (2004);
        F. Karsch,
        AIP Conf. Proc. {\bf 842}, 20 (2006).

\bibitem{Gavai:2005PRD}
        R. V. Gavai and S. Gupta,
        Phys. Rev. D {\bf 71}, 114014 (2005);
        {\bf 72}, 054006 (2005);
        {\bf 73}, 014004 (2006);
        Ph. de Forcrand and O. Philipsen,
        Nucl. Phys. {\bf B673}, 170 (2003).

\bibitem{Megias:2006PRD}
        E. Megias, E. Ruiz Arriola, and L.L. Salcedo,
        Phys. Rev. D {\bf 74}, 065005 (2006);
        Phys. Rev. D {\bf 74}, 114014 (2006);
        M. Ciminale, R. Gatto, N. D. Ippolito, G. Nardulli, and M. Ruggieri,
        Phys. Rev. D {\bf 77}, 054023 (2008).

\bibitem{Pisa1}
        R.D. Pisarski,
        Phys. Rev. D {\bf 62}, 111501(R) (2000);
        R.D. Pisarski,
        hep-ph/0203271.

\bibitem{Ratti:2005PRD}
        C. Ratti, M. A. Thaler, W. Weise,
        Phys. Rev. D {\bf 73}, 014019 (2006).

\bibitem{Muller}
        H.-M. Tsai and B. M\"{u}ller, 
        J. Phys. G: Nucl. Part. Phys. {\bf 36}, 075101 (2009).

\bibitem{Klevansky}
        P. Zhuang, J. H{\"{u}}fner, and S.P. Klevansky,
        Nucl. Phys. {\bf A576}, 525 (1994).

\bibitem{Buballa:2004PR}
        M. Buballa,
        Phys. Rep. {\bf 407}, 205 (2005).

\bibitem{Costa:2003PRC}
        P. Costa, M. C. Ruivo, C. A. de Sousa, and Y. L. Kalinovsky,
        Phys. Rev. C {\bf 70}, 025204 (2004).

\bibitem{Costa:2008PRD2}
        P. Costa, M.C. Ruivo, and C.A. de Sousa,
        Phys. Rev. D {\bf 77}, 096009 (2008).

\bibitem{Fukushima:2008PRD}
        K. Fukushima,
        Phys. Rev. D {\bf 77}, 114028 (2008).

\bibitem{Klevansky-review}
    		S.~P.~Klevansky,
    		Rev.\ Mod.\ Phys.\  {\bf 64}, 649 (1992).

\bibitem{Hansen:2007PRD}
        H. Hansen, W.  Alberico, A. Beraudo, A. Molinari, M. Nardi, and C. Ratti,
        Phys. Rev. D {\bf 75}, 065004 (2007).

\bibitem{kunihiro}
        T. Hatsuda and T. Kunihiro,
        Phys. Rep. {\bf 247}, 221 (1994).

\bibitem{fiolhais}
       	M. Fiolhais, J. da Provid\^encia, M. Rosina, and C. A. de Sousa,
        Phys. Rev. C {\bf 56}, 3311 (1997).

\bibitem{Buballa:1996NPA}
        M. Buballa,
        Nucl. Phys. A {\bf 611}, 393 (1996).

\bibitem{Costa:2008PRD3}
        P. Costa, C.A. de Sousa, M.C. Ruivo, and H. Hansen,
        Europhys. Lett. {\bf 86}, 31001 (2009).

\bibitem{Sasaki}
        C. Sasaki, B. Friman, and K. Redlich,
        Phys. Rev. D {\bf 75}, 054026 (2007);
        Phys. Rev. D {\bf 75}, 074013 (2007).

\bibitem{Kashiwa:2008PLB}
        K. Kashiwa, H. Kouno, M. Matsuzaki, and M. Yahiro,
        Phys. Lett. B {\bf 662}, 26 (2008).

\bibitem{Alb2009}
				P. Costa, M. C. Ruivo, C. A. de Sousa, H. Hansen, and W. M. Alberico, 
				Phys. Rev. D {\bf 79}, 116003 (2009).

\bibitem{Mish2000}
        I. N. Mishustin, L. M. Satarov, H. St\"{o}cker, and W. Greiner,
        Phys. Rev. C {\bf 62}, 034901 (2000).

\bibitem{Costa:2008PRD1}
        P. Costa, M.C. Ruivo, and C. A. de Sousa,
        Phys. Rev. D {\bf 77}, 096001 (2008).

\bibitem{Rajagopal:1999NPA}
        K. Rajagopal,
        Nucl. Phys. {\bf A661}, 150 (1999);
        J. Berges and K. Rajagopal,
        Nucl. Phys. {\bf B538}, 215 (1999).

\bibitem{Scavenius}
        O. Scavenius, A. Mocsy, I. N. Mishusti, and D. H. Rischke,
        Phys. Rev. C {\bf 64}, 045202 (2001).

\bibitem{Karsch2}
        F. Karsch,
        Lect. Notes Phys. {\bf 583}, 209 (2002);
        F. Karsch, E. Laermann, and A. Peikert,
        Nucl. Phys. {\bf B605}, 579 (2001).

\bibitem{lattice1}
        O. Philipsen,
        Eur. Phys. J. ST {\bf 152}, 29 (2007).

\bibitem{Ejiri:2006PRD}
        S. Ejiri, F. Karsch,E. Laermann, and C. Schmidt,
        Phys. Rev. D {\bf 73}, 054506 (2006).

\bibitem{Nonaka:2005PRC}
        C. Nonaka  and M. Asakawa,
        Phys. Rev. C {\bf 71}, 044904 (2005).

\bibitem{Kahara:2008}
        T. K\"{a}h\"{a}r\"{a} and K. Tuominen, 
        Phys. Rev. D {\bf 78}, 034015 (2008).

\bibitem{Subramanian}
        P. R. Subramanian, H. Stocker, and W. Greiner,
        Phys. Lett. B {\bf 173}, 468 (1986).

\bibitem{Stephanov:1999PRD}
        Stephanov M,  Rajagopal K and  Shuryak E, 
        1999 {\em Phys. Rev.} D {\bf 60} 114028

\bibitem{Fukushima} 
        K. Fukushima,
        Phys. Rev. D {\bf 79}, 074015 (2009);
        J. Phys. G: Nucl. Part. Phys. {\bf 36}, 064020 (2009).

\bibitem{Nakano} 
				E. Nakano, B.-J. Schaefer, B. Stokic, B. Friman, K. Redlich,
				arXiv:0907.1344 [hep-ph].         

\bibitem{Rossner:2007ik}
        S. Rossner, T. Hell, C. Ratti and W. Weise,
        Nucl. Phys. {\bf A814}, 118 (2008).

\bibitem{Griffiths:1970PR}
        R. B. Griffiths and J. Wheeler,
        Phys. Rev. A {\bf 2}, 1047 (1970).

\bibitem{lattice2}
        C.R. Allton, S. Ejiri, S.J. Hands, O. Kaczmarek, F. Karsch, E. Laermann, and C. Schmidt,
        Phys. Rev. D {\bf 68}, 014507 (2003).

\end{thebibliography}
\end{document}